%% file: main.tex
\documentclass[10pt,journal,compsoc]{IEEEtran}

\input{fei_shared}

\title{Fast Equivariant Imaging: Accelerating Unsupervised Learning and Model Adaptation via Inexact Splitting}

\author{Guixian~Xu,
        Jinglai~Li,
        and~Junqi~Tang%
\IEEEcompsocitemizethanks{\IEEEcompsocthanksitem G. Xu, J. Li, and J. Tang are with the School of Mathematics, University of Birmingham, Birmingham B15 2TT, United Kingdom.\protect\\
E-mail: \{gxx422@student.bham.ac.uk, j.li.10@bham.ac.uk, j.tang.2@bham.ac.uk\}.}%
}

\IEEEtitleabstractindextext{%
\begin{abstract}
In this work, we propose Fast Equivariant Imaging (FEI), a novel unsupervised learning framework to rapidly and efficiently train deep imaging networks without ground-truth data. FEI reformulates the EI objective through an inexact variable-splitting scheme, decoupling network training from an auxiliary restoration step implemented with a plug-and-play denoiser, this novel unsupervised scheme shows superior efficiency and performance compared to the standard Equivariant Imaging paradigm. In particular, our FEI schemes achieve an order-of-magnitude (10x) acceleration over standard EI on training U-Net for X-ray CT reconstruction and image inpainting, with improved generalization performance. Beyond offline training, the proposed scheme also enables efficient test-time adaptation of a pretrained model to individual samples, to secure further performance improvements. Extensive experiments show that the proposed approach provides a noticeable efficiency and performance gain over existing unsupervised methods and model adaptation techniques.
\end{abstract}

}

\begin{document}

\maketitle
\IEEEdisplaynontitleabstractindextext
\IEEEpeerreviewmaketitle

\IEEEraisesectionheading{\section{Introduction}}
Modern learning-based solvers have emerged as a powerful paradigm, typically trained in a supervised manner using large external datasets that contain paired ground truth (GT) images and measurements~\cite{xin2016maximal, zhang2018ista}. However, in certain domains, such as medical imaging, it is often prohibitively expensive or even impossible to obtain sufficient GT data for supervised learning~\cite{belthangady2019applications}. Consequently, models trained on such limited data are susceptible to distribution bias, which can lead to poor generalization performance on unseen test samples.

Recently, there has been an increased interest in developing various methodologies to address potential biases and poor generalization due to limited training data. Approaches based on \textit{self-supervised internal learning}, exemplified by Deep Image Prior (DIP)~\cite{ulyanov2018deep}, leverage the inherent inductive bias of network architectures. DIP exploits the observation that untrained CNNs converge to regular structures faster than to noise during optimization. Although this approach allows for the precise exploitation of instance-specific characteristics, it incurs high computational costs as it necessitates optimizing the network from scratch for each test sample. In contrast, \textit{self-supervised external learning} addresses the absence of ground truth data by training a shared model on a dataset of measurements~\cite{quan2022dual, chen2021equivariant, chen2022robust}. However, these methods generally suffer from the same limitation as supervised learning: susceptibility to poor generalization when test samples deviate from the training distribution.

Another significant line of research encompasses \textit{unsupervised meta-learning} and \textit{test-time adaptation} (TTA). Meta-learning aims to train a model capable of rapid adaptation to new tasks with minimal data. However, the majority of meta-learning approaches are supervised~\cite{chi2021test, park2020fast}, relying on GT images to fine-tune pre-trained models. \cite{wang2024test} proposed a hybrid strategy that involves supervised meta-training and self-supervised meta-testing for medical imaging, utilizing an equivariant-based self-supervised loss. In contrast, purely unsupervised meta-learning for inverse imaging problems remains underexplored. A notable exception is \cite{qin2023ground}, who proposed an unsupervised meta-learning framework for medical imaging employing a SURE-based loss.

Parallel to meta-learning, TTA seeks to adjust a pre-trained model during inference to align it with the test data distribution, thereby mitigating performance degradation. \cite{darestani2022test} demonstrated promising results by enforcing a data consistency (DC) loss during both the training phase and TTA. Specifically, they employ a self-supervised approach where reconstructed results are re-sampled to ensure consistency with the input measurements. However, relying solely on DC loss for TTA is often insufficient to learn robust self-supervised features~\cite{chen2021equivariant}, frequently resulting in suboptimal or even degraded reconstruction quality. Although \cite{zhao2024test} integrated self-supervised loss into deep unfolding networks via a lightweight adaptive layer, the considerable depth of the unrolled architecture results in slow adaptation speeds and limited robustness against significant distribution shifts.

Equivariant Imaging (EI) unifies \textit{self-supervised external learning} and \textit{test-time adaptation} through a natural objective based on equivariance consistency~\cite{wang2024test, chen2021equivariant}. This objective enforces consistency between the reconstructed images and the transformations applied to the input. Using the inherent symmetries of the imaging process and the physical measurement operator, EI facilitates more stable and effective learning and adaptation. However, EI faces significant computational and optimization challenges. Training is notably slower because each iteration requires multiple model evaluations, resulting in high computational overhead. Furthermore, the equivariant loss is fully effective only when the estimator achieves a near-perfect reconstruction. Consequently, this supervision signal is weak during the early stages of training, often leading to slow convergence.

In this work, we introduce Fast Equivariant Imaging (FEI), an efficient unsupervised training and adaptation framework that makes EI practical for high-dimensional tasks. FEI is based on a variable-splitting reformulation of the original EI objective that converts the single computationally expensive training problem into a sequence of simpler subproblems. Sepcifically, FEI alternates between two stages: a \textbf{latent-reconstruction step} that refines an auxiliary estimate of the ground truth, and a \textbf{pseudo-supervision step} that enforces equivariance on the network using the refined latent. From an optimization perspective, variable splitting yields an equivalent constrained formulation that can be solved using Lagrangian methods~\cite{afonso2010fast}. { While standard variable splitting~\cite{afonso2010fast} requires exact subproblem minimization, recent advances in inexact optimization \cite{devolder2014first} motivate interpreting relaxed schemes through the size of their approximation errors. Hence, we introduce a task-decoupled splitting strategy: the latent-reconstruction step focuses on solving the inverse problem via measurement fidelity and image priors, while the equivariance constraint is enforced as a regularization on the network parameters during the pseudo-supervision step. This design avoids the computationally expensive evaluation of equivariance gradients w.r.t. the latent variable, leading to substantial reductions in runtime and memory.} Although we focus only on accelerating EI in this work, the same algorithmic structure can be easily extended to more recent variants of EI such as Robust Equivariant Imaging (REI) \cite{chen2022robust}, Multi-Operator Imaging (MOI) \cite{tachella2022unsupervised}, and our recently proposed Sketched Equivariant Imaging (SkEI) \cite{xu2024sketched}, which utilizes dimensionality reduction techniques for acceleration.

We summarize the specific technical contributions of this work as follows:
\begin{itemize}
    \item We propose an efficient unsupervised training paradigm FEI which conceptually splits the original brute-force EI training into two alternating steps: a Latent-Reconstruction Step and a Pseudo-Supervision Step. We demonstrate numerically that such a splitting can massively accelerate the unsupervised training of deep imaging networks.
    \item Under the FEI splitting framework, we propose two novel optimization schemes tailored for efficiently implementing FEI, based on a new integration of adaptive gradient methods (such as Adam) with inexact HQS-inspired and Linearized ADMM-inspired splitting. We utilized the gradient history of the subproblems from previous iterations to update the network parameters and promote efficient optimization in practice.
    \item Building upon FEI, we also propose the first unsupervised training scheme PnP-FEI which utilizes pretrained denoisers for further numerical improvements. Thanks to the splitting provided by FEI, we can effectively utilize pretrained image-domain priors such as deep denoiser within our framework, in the Latent-Reconstruction Step. This is the first paradigm which utilizes both the primal domain (image) prior and the dual domain (measurement) prior for unsupervised training of deep imaging networks.

    \item In addition, we demonstrate that the proposed scheme enables efficient test-time adaptation of a pretrained model to individual samples to secure further performance improvements and robustness towards distribution shifts. 
\end{itemize}


\section{Related Work}
\subsection{Variable Splitting}
The variable splitting strategy, originally formalized by~\cite{afonso2010fast}, decouples the data fidelity term from non-smooth regularization components. This strategy has been widely adopted to guide deep unrolling frameworks in supervised learning, such as~\cite{zhang2020deep, chan2016plug, yang2018admm}. Beyond supervised learning, variable splitting has proven critical for stabilizing deep image prior (DIP), approaches like DeepRED~\cite{mataev2019deepred} utilize splitting to impose explicit regularization (e.g., RED) on the network output, effectively mitigating the overfitting and instability inherent to original DIP optimization. Furthermore, this modularity extends to self-supervised learning, where architectures like DURED-Net~\cite{huang2023self} leverage variable splitting to isolate data consistency from denoising tasks, enabling robust physics-guided reconstruction via measurement splitting across multiple undersampled operators. To the best of our knowledge, this work is the first to extend variable splitting to the more challenging unsupervised setting involving single-operator inverse problems. This domain notably encompasses critical applications such as sparse-view CT and accelerated MRI with fixed sampling masks.

\subsection{Equivariant Imaging} Among unsupervised approaches for training deep imaging networks, the Equivariant Imaging (EI) framework~\cite{chen2021equivariant,chen2022robust} pioneered the explicit recovery of information beyond the range space of the physical operator by exploiting the inherent symmetries of imaging systems. EI posits that the distribution of clean images is invariant under specific geometric transformations, such as translations, rotations, and flips. Accordingly, it introduces a training loss that enforces equivariance throughout the measurement-reconstruction process, thereby constraining the set of learnable models. This unsupervised paradigm has recently achieved state-of-the-art performance in a diverse array of inverse problems~\cite{wang2024perspective, terris2024equivariant}. However, one of the main challenges of these schemes is the computational cost of training due to slow convergence and large computational complexity between training iterations, particularly for large-scale inverse problems and advanced architectures such as unrolling \cite{monga2021algorithm,zhou2024deep}. In our recent work of Sketched EI \cite{xu2024sketched} we partially addressed the computational complexity issue for inverse problems on which dimensionality reduction strategies can be applied.

\subsection{Test-Time Adaptation} TTA adapts a pre-trained model during the inference phase without accessing the original source data. The primary objective is to align the model with the specific distribution of the test data, thus enhancing performance in the target domain. By operating in a source-free manner, TTA effectively addresses concerns regarding data privacy and availability. To compensate for the absence of ground truth (GT) supervision during adaptation, various self-supervised loss functions have been proposed. For instance, MetaCS~\cite{qin2023ground} introduced a self-supervised loss based on Stein's Unbiased Risk Estimator (SURE) for unsupervised meta-learning. Similarly, DDSSL~\cite{quan2022dual} developed a self-supervised dual-domain loss specifically tailored for TTA in medical imaging contexts. However, while these approaches mitigate the reliance on GT images, their performance generally remains inferior to that of fully supervised methods.

\section{Background}
\subsection{Problem setup}
In this work, we address linear inverse imaging problems modeled as:
\begin{equation}
    \mathbf{y} \approx \mathbf{A} \mathbf{x}^\dagger,
\end{equation}
where $\mathbf{x}^\dagger \in \mathbb{R}^n$ denotes the unknown signal to be estimated and $\mathbf{y} \in \mathbb{R}^m$ represents the noisy measurements. The forward operator is given by $\mathbf{A} \in \mathbb{R}^{m \times n}$, often representing an ill-posed setting where $m \ll n$. In this context, a Maximum A Posteriori (MAP) estimator for $\mathbf{x}^\dagger$ can be formulated as the solution to the following optimization problem:
\begin{equation}
    \mathbf{x}^\star = \mathop{\arg \min}\limits_{\mathbf{x}} \{f_\mr{mc}(\mathbf{Ax}, \mathbf{y}) + R(\mathbf{x})\},
\end{equation}
 where $f_\mr{mc}(\mathbf{Ax}, \mathbf{y})$ is the measurement consistency (MC) term such as $\ell_1$ loss, $\ell_2$ loss and SURE-loss \cite{chen2022robust}, $R(\mathbf{x})$ is the prior or regularizer. Typically, learning based methods seek to train a network
 \begin{equation}
    \Ft( \mathbf{y}) \rightarrow \mathbf{x}
\end{equation}
approximating the solution of the aforementioned optimization problem. The network $\Ft$ can be trained using supervised or unsupervised strategies. Our focus lies on the latter scenario, where training is based solely on measurement data without access to ground-truth images. This is achieved primarily by enforcing measurement consistency, often complemented by additional regularization or self-supervised objectives.

\subsection{Equivariant Imaging}
To enforce data fidelity, we formulate the Measurement Consistency (MC) objective as:
\begin{equation}
    \min_{\theta}\mathbb{E}_{\mathbf{y} \in \mathcal{Y}} f_\mr{mc} ( \mathbf{A} \Ft(\mathbf{y}) , \mathbf{y}).
    \tag{MC}
\end{equation}
However, relying solely on the MC constraint is insufficient to uniquely determine the inverse mapping. As noted in~\cite{chen2021equivariant}, this objective does not recover signal components lying within the null space of the forward operator $\mathbf{A}$. 

To recover information lost during the forward process, \cite{chen2021equivariant} leverages prior information about the inherent symmetries of the signal set. Specifically, the distribution of plausible images is assumed to be invariant under certain groups of transformations (e.g., shifts, rotations). This symmetry implies that the composition $\mathcal{F}_\theta \circ \mathbf{A}$ should be equivariant to the transformation $\mathrm{T}_g$:
\begin{equation}
    \Ft (\mathbf{A} \mathrm{T}_g \mathbf{x}^\dagger) = \mathrm{T}_g \Ft(\mathbf{A}\mathbf{x}^\dagger)
\end{equation}
where $\mathrm{T}_g \in \mathbb{R}^{n \times n}$ denotes a unitary matrix representation of the group element $\mathcal{G}$. This equivariant regularization leads to the following unsupervised learning paradigm:
\begin{equation}\label{eq: EI}
    \min_{\theta} \mathbb{E}_{\mathbf{y} \in \mathcal{Y}, g \sim \mathcal{G}} \{f_\mr{mc} (\mathbf{y} , \mathbf{A} \Ft(\mathbf{y}) ) + \alpha \Vert \mathrm{T}_g \Ft(\mathbf{y}) - \Ft(\mathbf{A} \mathrm{T}_g \Ft(\mathbf{y})) \Vert_2^2 \},
\end{equation}
where $\alpha > 0$ is a trade-off coefficient. While EI has recently achieved state-of-the-art performance for various inverse problems~\cite{chen2021equivariant,chen2022robust, terris2024equivariant}, its effectiveness relies on the condition $\mathrm{T}_g (\mathcal{F}_\theta \circ \mathbf{A})\mathbf{x}^\dagger \approx (\mathcal{F}_\theta \circ \mathbf{A})\mathrm{T}_g \mathbf{x}^\dagger$. This relationship implies that the equivariant loss provides most meaningful gradients primarily when the estimator is already capable of near-perfect reconstruction ($\mathcal{F}_\theta (\mathbf{y}) \approx \mathbf{x}^\dagger$). However, this condition is rarely met during the early stages of training, resulting in an inefficient optimization trajectory.

\subsection{Variable Splitting and
{ Inexact} Optimization}
One strategy to address the limitations of EI is to employ variable splitting~\cite{afonso2010fast}. Variable splitting is a standard technique used to reformulate an unconstrained optimization problem of the form:
\begin{equation}\label{eq: vs-un}
    \min_{\mathbf{u} \in \mathbb{R}^n} f_1(\mathbf{u}) + f_2(g(\mathbf{u}))
\end{equation}
where $g: \mathbb{R}^n \to \mathbb{R}^d$. We introduce an auxiliary variable $\mathbf{v} \in \mathbb{R}^d$ to decouple the argument of $f_2$, imposing the constraint $g(\mathbf{u})=\mathbf{v}$. This yields the equivalent constrained formulation:
\begin{equation}\label{eq: vs-con}
    \min_{\mathbf{u} \in \mathbb{R}^n, \mathbf{v} \in \mathbb{R}^d} f_1 (\mathbf{u}) + f_2(\mathbf{v}), \quad \text{subject \ to} \quad g(\mathbf{u}) = \mathbf{v}.
\end{equation}
Problem~\eqref{eq: vs-con} is equivalent to the unconstrained problem~\eqref{eq: vs-un} over the feasible set $\{ (\mathbf{u}, \mathbf{v}) : g(\mathbf{u}) = \mathbf{v} \}$. The key idea is that solving the decoupled constrained problem~\eqref{eq: vs-con} is often computationally more tractable than solving its unconstrained counterpart~\eqref{eq: vs-un}, particularly when splitting allows for the use of efficient solvers for the sub-problems.

{ Standard splitting methods typically require the exact minimization of subproblems to guaranty convergence. However, in large-scale imaging tasks, exact solutions are often computationally intractable. Inexact splitting methods relax this requirement, allowing subproblems to be solved approximately (e.g., via limited gradient steps or by omitting complex regularization terms), and their approximation errors can often be analyzed explicitly~\cite{devolder2014first, yang2013linearized, robini2018inexact, robini2016inexact}. In this work, we use this viewpoint only to motivate a residual-controlled interpretation of the relaxed FEI splitting.}

\section{Method}
\begin{figure}[h!]
    \centering
    \includegraphics[width=0.6\linewidth]{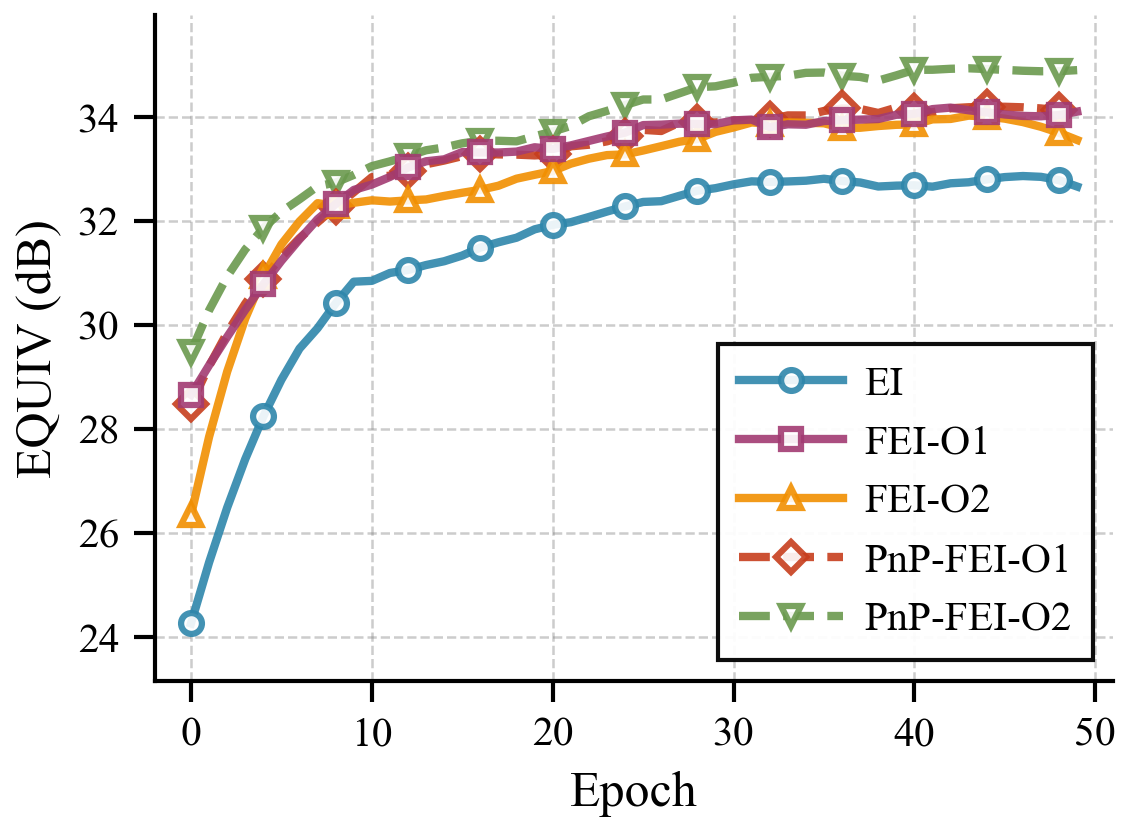}
    \caption{Evolution of the EQUIV metric during the early stages of training
    (see Sec.~\ref{sec: num_exp} for the definition). Since EQUIV is a logarithmic
    measure of the equivariance residual, a higher value indicates a smaller residual.
    The plot provides empirical support for the mechanism discussed above: the FEI
    latent-reconstruction step leads to a more effective equivariance signal during
    early training.}
    \label{fig:equiv-comp}
\end{figure}

In this section, we present our fast Equivariant Imaging (FEI) framework, which is based on splitting the original objective of EI into multiple subproblems. The original EI is enforced $\mathrm{T}_g \Ft(\mathbf{y}) \approx \Ft(\mathbf{A}\mathrm{T}_g \Ft(\mathbf{y}))$. Crucially, the underlying assumption $\mathrm{T}_g (\Ft \circ \mathbf{A})\mathbf{x}^\dagger = (\Ft \circ \mathbf{A})\mathrm{T}_g \mathbf{x}^\dagger$ holds strictly at the ground truth $\mathbf{x}^\dagger$. This implies that the equivariant loss is most effective when the estimator achieves near-perfect reconstruction, i.e., $\Ft(\mathbf{y}) \approx \mathbf{x}^\dagger$. Consequently, improving reconstruction quality facilitates the enforcement of equivariance. Figure~\ref{fig:equiv-comp} provides an empirical illustration consistent with this interpretation, showing that FEI improves the EQUIV metric more rapidly than vanilla EI during the early stage of training. Based on this insight, we propose decomposing the original optimization problem into multiple steps:  \textbf{Latent-reconstruction} and \textbf{Pseudo-supervision}. 

Denote the Latent-reconstruction mapping as 
\[
M_L(\mathbf{y}, \mathcal{F}_{\theta_k}) \rightarrow \mathbf{u}_{k+1},
\] 
which takes current network output at a data sample $\mathbf{y}$ and form a latent image $\mathbf{u}_{k+1}$. We then also denote the Pseudo-supervision mapping as 
\[
M_P(\mathbf{u}_{k+1}, \mathcal{F}_{\theta_k}) \rightarrow \theta_{k+1}
\]
which takes the latent image to perform the update on network parameters joint-force with EI-regularization. Then we can describe our generic FEI framework here:
  \begin{eqnarray*}
 && \mathrm{\textbf{FEI generic framework}}\\&& - \mathrm{Initialize}\ \theta_0\\
 &&\mathrm{For} \ \ \ k = 0, 1, 2,...,  K\\
&&\left\lfloor
\begin{array}{l}
\mathrm{Sample\ } \mathbf{y} \in \mathcal{Y}, g \sim \mathcal{G} \\
\mathbf{u}_{k+1} \leftarrow M_L(\mathbf{y}, \mathcal{F}_{\theta_k})  \ \ \ \ \ \ \ \ \ \ \ \ \ \ \mathrm{\it\color{purple} Latent\ reconstruction}\\
\theta_{k+1} \leftarrow M_P(\mathbf{u}_{k+1}, \mathcal{F}_{\theta_k})  \ \ \ \ \ \ \ \  \mathrm{\it\color{purple} Pseudo\ supervision}
\end{array}
\right.
\end{eqnarray*}

This leads to (at least) two options for FEI algorithmically, which we describe as HQS-inspired and ADMM-inspired relaxed splitting schemes. We build our numerical training algorithms for implementing FEI via modifying both HQS and ADMM splitting and tailoring them for unsupervised training of deep imaging networks. Crucially, addressing the equivariance constraint $\alpha \|\mathrm{T}_g \mathbf{u} - \mathcal{F}_\theta(\mathbf{A} \mathrm{T}_g \mathbf{u})\|^2$ strictly within the latent-reconstruction step (updating $\mathbf{u}$) poses a significant computational bottleneck. It would require backpropagation through the network $\mathcal{F}_\theta$ at every inner iteration of the solver, thus negating the efficiency gains of variable splitting. To avoid this, we propose a relaxed splitting scheme that decouples equivariance enforcement from latent variable update. We specifically assign the role of maintaining equivariance to the network parameter update step (Pseudo-supervision), while the latent step focuses on enforcing measurement consistency and signal priors.

\subsection{Residual-controlled inexact splitting of FEI}
\label{subsec:fei_theory}

We now give a residual-controlled inexact splitting interpretation of the relaxed FEI update.
The practical implementation uses Adam for the network update and either an accelerated or linearized update for the latent-reconstruction step.
Our goal here is: we isolate the structural error introduced when the latent step omits the equivariance term, and show that this error is controlled by the same equivariance residual minimized in the pseudo-supervision step.

This viewpoint is related to inexact block-coordinate methods for nonconvex optimization~\cite{bolte2014proximal,yang2019inexact}.
FEI has a natural two-block structure with respect to the latent image \(\mathbf{u}\) and the network parameter \(\theta\).
However, unlike an exact split update, the latent-reconstruction step is intentionally performed on a partial objective that excludes the equivariance term.
This relaxation avoids differentiating the equivariance loss through the network inside the latent solver, which is the main computational bottleneck in the exact split update.

For notational simplicity, we state the analysis for a fixed measurement \(\mathbf{y}\) and a fixed transformation \(\mathrm{T}_g\).
The finite-sample full-batch objective can be handled by summing or averaging the same objective over training samples and sampled transformations.
Define the split FEI objective
\begin{equation}
\label{eq:split_fei_objective}
\mathcal L_{\lambda,\alpha}(\mathbf{u},\theta)
=
f_{\rm mc}(\mathbf{A}\mathbf{u},\mathbf{y})
+
\frac{\lambda}{2}\|\mathbf{u}-\Ft(\mathbf{y})\|_2^2
+
\alpha\|\mathrm{T}_g \mathbf{u}-\Ft(\mathbf{A}\mathrm{T}_g \mathbf{u})\|_2^2 .
\end{equation}
Here, \(\mathbf{u}\) denotes the latent reconstruction variable and \(\theta\) denotes the network parameters.
The first two terms correspond to the latent-reconstruction step, while the last term corresponds to the equivariance regularization used in the pseudo-supervision step.

We write
\begin{equation}
\label{eq:split_decomposition}
\mathcal L_{\lambda,\alpha}(\mathbf{u},\theta)
=
\Phi(\mathbf{u},\theta)+\Psi(\mathbf{u},\theta),
\end{equation}
where
\begin{equation}
\label{eq:phi_definition}
\Phi(\mathbf{u},\theta)
=
f_{\rm mc}(\mathbf{A}\mathbf{u},\mathbf{y})
+
\frac{\lambda}{2}\|\mathbf{u}-\Ft(\mathbf{y})\|_2^2
\end{equation}
and
\begin{equation}
\label{eq:psi_definition}
\Psi(\mathbf{u},\theta)
=
\alpha\|\mathrm{T}_g \mathbf{u}-\Ft(\mathbf{A}\mathrm{T}_g \mathbf{u})\|_2^2 .
\end{equation}

A full \(\mathbf{u}\)-block update for \(\mathcal L_{\lambda,\alpha}\) would use the partial gradient
\(\nabla_{\mathbf{u}}\mathcal L_{\lambda,\alpha}(\mathbf{u},\theta)\).
In contrast, the FEI latent-reconstruction step updates \(\mathbf{u}\) using only the partial objective \(\Phi\):
\begin{equation}
\label{eq:partial_u_update_main}
\mathbf{u}_{k+1}
\approx
\mathbf{u}_k-\eta_u \nabla_{\mathbf{u}} \Phi(\mathbf{u}_k,\theta_k),
\end{equation}
or a finite number of accelerated or linearized steps for approximately reducing \(\Phi(\cdot,\theta_k)\).
Equivalently,
\begin{equation}
\label{eq:structured_error_main}
\nabla_{\mathbf{u}}\Phi(\mathbf{u}_k,\theta_k)
=
\nabla_{\mathbf{u}}\mathcal L_{\lambda,\alpha}(\mathbf{u}_k,\theta_k)
-
e_k^u,
\qquad
e_k^u:=\nabla_{\mathbf{u}}\Psi(\mathbf{u}_k,\theta_k).
\end{equation}
Thus, the relaxed latent-reconstruction step can be viewed as an inexact \(\mathbf{u}\)-block update for the full split objective, where the structured error is
\begin{equation}
\label{eq:omitted_gradient_error_main}
e(\mathbf{u},\theta):=\nabla_{\mathbf{u}}\Psi(\mathbf{u},\theta).
\end{equation}

The following proposition shows that this inexactness is controlled by the same equivariance residual that the pseudo-supervision step explicitly reduces.
This gives a direct bridge between the computational relaxation used in FEI and a residual-controlled notion of inexact splitting.

\begin{proposition}[Omitted-gradient error controlled by equivariance residual]
\label{lem:main_omitted_gradient_residual}
Suppose \(\mathrm{T}_g\) and \(\mathbf{A}\) are bounded linear operators, and suppose \(\Ft\) is differentiable with respect to its image input on the relevant region with
\[
\|J_{\mathbf{x}}\Ft(\mathbf{x})\|\leq C_x .
\]
Let
\[
\Psi(\mathbf{u},\theta)
=
\alpha\|\mathrm{T}_g \mathbf{u}-\Ft(\mathbf{A}\mathrm{T}_g \mathbf{u})\|_2^2
\]
and define the equivariance residual
\begin{equation}
\label{eq:app_eq_residual_definition}
R_{\rm eq}(\mathbf{u},\theta)
:=
\mathrm{T}_g \mathbf{u}-\Ft(\mathbf{A}\mathrm{T}_g \mathbf{u}).
\end{equation}
Then the omitted-gradient error
\[
e(\mathbf{u},\theta):=\nabla_{\mathbf{u}}\Psi(\mathbf{u},\theta)
\]
satisfies
\begin{equation}
\label{eq:e_residual_bound_general}
\|e(\mathbf{u},\theta)\|_2
\leq
2\alpha \|\mathrm{T}_g\|
\left(
1+\|\mathbf{A}\|\,C_x
\right)
\|R_{\rm eq}(\mathbf{u},\theta)\|_2 .
\end{equation}
In particular, if \(\mathrm{T}_g\) is an isometry, then
\begin{equation}
\label{eq:e_residual_bound_isometry}
\|e(\mathbf{u},\theta)\|_2
\leq
2\alpha
\left(
1+\|\mathbf{A}\|\,C_x
\right)
\|\mathrm{T}_g \mathbf{u}-\Ft(\mathbf{A}\mathrm{T}_g \mathbf{u})\|_2 .
\end{equation}
\end{proposition}

\begin{proof}
The proof is given in Appendix~\ref{app:omitted_gradient_proof}.
\end{proof}

For isometric transformations \(\mathrm{T}_g\), the factor \(\|\mathrm{T}_g\|\) equals one.
Hence, the error introduced by the relaxed latent update is controlled by the same equivariance residual that FEI reduces in the pseudo-supervision step. 

Moreover, the splitting structure of FEI naturally allows the incorporation of image-domain priors in the latent-reconstruction step.
This motivates the PnP-FEI extension, where a plug-and-play denoiser is used as an additional primal-domain regularizer, while EI continues to provide measurement-domain equivariant regularization.

\subsection{Fast Equivariant Imaging -- Option 1 ({ HQS-inspired relaxed splitting})}
We first reformulate~\eqref{eq: EI} by substituting $\mathbf{u} = \Ft(\mathbf{y})$ as follows
\begin{equation}\label{eq: fei}
    \begin{aligned}
    \min_{\theta, \mathbf{u}} \mathbb{E}_{\mathbf{y} \in \mathcal{Y},\, g \sim \mathcal{G}} \Bigl\{
    &f_\mr{mc}(\mathbf{y}, \mathbf{Au}) + \alpha \Vert \mathrm{T}_g \mathbf{u} - \Ft(\mathbf{A} \mathrm{T}_g \mathbf{u}) \Vert_2^2
    \Bigr\}, \\
    &\text{s.t.}\quad \mathbf{u} = \Ft(\mathbf{y}).
    \end{aligned}
\end{equation}
To solve this constraint problem with the choice of Euclidean norm of $f$, we first rewrite the objective with penalty parameter $\lambda$,
\begin{equation}\label{eq: hqs}
    \begin{aligned}
    \mathcal{L}_{\lambda, \alpha} (\mathbf{u}, \theta)
    &= \mathbb{E}_{\mathbf{y} \in \mathcal{Y},\, g \sim \mathcal{G}} \Bigl\{
    f_\mr{mc}(\mathbf{y}, \mathbf{Au}) + \frac{\lambda}{2} \Vert \mathbf{u} - \Ft(\mathbf{y}) \Vert_2^2 \\
    &\qquad\qquad\qquad\qquad + \alpha \Vert \mathrm{T}_g \mathbf{u} - \Ft(\mathbf{A} \mathrm{T}_g \mathbf{u}) \Vert_2^2
    \Bigr\}.
    \end{aligned}
\end{equation}
We then address problem~\eqref{eq: hqs} with HQS-inspired splitting algorithm due to its simplicity. Note that a very large $\lambda$ will force $\mathbf{u}$ to be approximately equal to $\Ft(\mathbf{y})$. Usually, $\lambda$ varies in a non-descending order during the following iterative solution
\begin{align}
    \mathbf{u}_{t+1} &\approx \arg \min_{\mathbf{u}} \mathcal{L}_{\lambda, \alpha}(\mathbf{u}, \theta_t)\label{eq: sub1}; \ \ \rightarrow \mathrm{\it\color{purple} Latent\ reconstruction\ step} \\
    \theta_{t+1} &\approx \arg \min_{\theta} \mathcal{L}_{\lambda, \alpha}(\mathbf{u}_{t+1}, \theta); \ \ \rightarrow \mathrm{\it\color{purple} Pseudo\ supervision\ step} \label{eq: sub2}
\end{align}
{To ensure the convexity and computational tractability of the $\mathbf{u}$-update, we adopt an inexact Half-Quadratic Splitting (HQS) strategy where the equivariance regularization is deferred to the network update step ($\theta$).} Consequently, the latent reconstruction subproblem~\eqref{eq: sub1} simplifies to refine the estimate $\mathbf{u}$:
\begin{equation}\label{eq: sub11}
\min_{\mathbf{u}} { \mathbb{E}_{\mathbf{y} \in \mathcal{Y}}} \left\{ f_\mathrm{mc}(\mathbf{Au}, \mathbf{y}) + \frac{\lambda}{2} || \mathbf{u} - \mathcal{F}_\theta(\mathbf{y}) ||_2^2 \right\}.
\end{equation}
We propose solving~\eqref{eq: sub11} via Nesterov Accelerated Gradient (NAG)~\cite{nesterov1983method, nesterov2007gradient}. Of course, one could also use other optimizers, such as stochastic gradient methods \cite{ehrhardt2024guide} (if the objective can be minibatched) or quasi-Newton methods \cite{tan2024provably} for this step.

However, the Lagrangian with respect to $\theta$ is nonconvex, which means the minimization over $\theta$ is often intractable. Here we bypass it by calling backpropagating over $\theta$ and taking one gradient descent step using an adaptive gradient method (Adam by default). The complete algorithm for Lagrangian-based fast equivariant imaging can be found in Algorithm~\ref{alg:LEI}. Note that in each iteration we typically just run one iteration of Adam for the Pseudo-supervision step, while the gradient and momentum history of Adam is inherited from previous Pseudo-supervision steps.

\begin{algorithm}[h!]
    \caption{Fast Equivariant Imaging (FEI-Option 1)}
    \label{alg:LEI}
    \begin{algorithmic}[1] 
    \REQUIRE 

    \begin{tabular}[t]{@{}p{0.45\linewidth}@{}p{0.45\linewidth}@{}}
    $\bullet$ $\mathcal{Y}$ - Observations (training dataset) & $\bullet$ $\mathbf{u}_0$ - initial NAG point \\
    
    $\bullet$ $\lambda$ - Penalty parameter & $\bullet$ $\mathbf{v}_0$ - initial NAG velocity point \\
    
    $\bullet$ $\alpha$ - EI regularization parameter & $\bullet$ $\beta$ - NAG momentum factor \\
    
    $\bullet$ $N$ - number of epochs & $\bullet$ $\eta$ - NAG step size \\
    
    $\bullet$ $J$ - number of NAG iterations & $\bullet$ $\mathcal{G}$ - Group of transformations \\
    \end{tabular}
        
        \STATE \textbf{Initialize:} Neural network $\Ft$, $\mathbf{u}_0 = \mathbf{A}^\dagger \mathbf{y}$, $\mathbf{v}_0 = \mathbf{0}$, $k=0$;
        
        \FOR{$i=0, \dots, N-1$}
            \FOR{$\mathbf{y} \in \mathcal{Y}$}
                \STATE Sample $\mathrm{T}_g$ where $g \sim \mathcal{G}$;
                \STATE Compute $\mathbf{x}_0 = \mathcal{F}_{\theta_k}(\mathbf{y})$, $\mathbf{u}_k = \mathbf{x}_0$;
                \STATE Set $f(\mathbf{u}) = f_\mr{mc}(\mathbf{y}, \mathbf{Au}) + \frac{\lambda}{2} \Vert \mathbf{u} - \mathcal{F}_{\theta_k}(\mathbf{y}) \Vert_2^2$;
                \STATE Update $\mathbf{u}_{k+1} = \mathcal{A}(f, J, \mathbf{u}_k, \mathbf{v}_k, \beta, \eta)$ via NAG algorithm;
                \STATE Let $\mathbf{x}_1 = \mathbf{u}_{k+1}$;
                \STATE Transform $\mathbf{x}_2 = \mathrm{T}_g(\mathbf{x}_1)$;
                \STATE Compute $\mathbf{x}_3 = \Ft(\mathbf{A} \mathbf{x}_2)$;
                \STATE Update $\theta_{k+1}$ via progressively minimizing loss $\mathcal{L} = \mathbb{E}_{g \sim \mathcal{G}} \{ \Vert \mathcal{F}_{\theta}(\mathbf{y}) - \mathbf{x}_1 \Vert^2_2 + \alpha \Vert \mathbf{x}_2 - \mathbf{x}_3 \Vert^2_2 \}$ using a gradient-based optimizer (such as Adam) in each iteration; 
                \STATE $k \leftarrow k+1$
            \ENDFOR
        \ENDFOR
        
        \ENSURE $\mathcal{F}_{\theta^\star}$
    \end{algorithmic}
\end{algorithm}

\subsection{Fast Equivariant Imaging --- Option 2 ({ Linearized ADMM-inspired relaxed splitting})}
We now present the second option of splitting style for FEI, which is based on modifying the ADMM splitting and tailoring it to unsupervised learning for imaging. First, starting with~\eqref{eq: fei}, we turn the constraint into a penalty using the scaled Augmented Lagrangian (AL)~\cite{afonso2010fast},
\begin{equation}\label{eq: admm-ei}
   \begin{aligned}
   \mathcal{L}_{\lambda, \alpha}(\mathbf{u}, \theta, \mathbf{L})
   := \mathbb{E}_{\mathbf{y} \in \mathcal{Y},\, g \in \mathcal{G}} \Bigl\{
   &f_{\textbf{mc}}(\mathbf{y}, \mathbf{A}\mathbf{u}) + \alpha \Vert \mathrm{T}_g \mathbf{u} - \Ft( \mathbf{A}\mathrm{T}_g \mathbf{u}) \Vert_2^2 \\
   &+ \frac{\lambda}{2} \Vert \mathbf{u} - \Ft (\mathbf{y}) + \mathbf{L} \Vert_2^2
   \Bigr\}.
   \end{aligned}
\end{equation}
for a penalty weight $\lambda >0$. Applied to~\eqref{eq: admm-ei}, every iteration would minimize the Augmented Lagrange with respect to $\mathbf{u}$, then with respect to $\theta$, and then update the dual variable $\mathbf{L}$, which are presented as follows:
\begin{equation}\label{eq: admm-sub}
    \begin{aligned}
        \mathbf{u}_{k+1} &\approx \mathop{\arg\min}_{\mathbf{u}} \mathcal{L}_{\lambda, \alpha}(\mathbf{u}, \theta_k, \mathbf{L}_k),\ \ \rightarrow \mathrm{\it\color{purple} Latent\ reconstruction\ step} \\
        \theta_{k+1} &\approx \mathop{\arg\min}_{\theta} \mathcal{L}_{\lambda, \alpha}(\mathbf{u}_{k+1}, \theta, \mathbf{L}_k),  \ \ \rightarrow \mathrm{\it\color{purple} Pseudo\ supervision\ step} \\
        \mathbf{L}_{k+1} &= \mathbf{L}_k +  \mathbf{u}_{k+1}-\mathcal{F}_{\theta_{k+1}}(\mathbf{y}).
    \end{aligned}
\end{equation}
then take only one descent step in both $\mathbf{u}$ and $\theta$. Specifically, fixing $\theta_k, \mathbf{L}_k$, the update of $\mathbf{u}_{k+1}$ can be performed by one step gradient descent { on the measurement consistency and Lagrangian terms. Consistent with our inexact splitting formulation, the equivariance regularization is reserved for the subsequent $\theta$-update step to avoid nested gradient evaluations.} Then fixing $\mathbf{u}_{k+1}, \mathbf{L}_k$, the update of $\theta$ is done by calling backpropagating and taking one adaptive gradient descent (Adam by default) step $\theta$. Note that in each iteration we typically just run one iteration of Adam for the Pseudo-supervision step, while the gradient and momentum history of Adam is inherited from previous Pseudo-supervision steps. As for the dual variable vector $\mathbf{L}$, its update is much easier, given by
\begin{equation}
    \mathbf{L}_{k+1} = \mathbf{L}_k + \mathbf{u}_{k+1} - \mathcal{F}_{\theta_{k+1}}(\mathbf{y})
\end{equation}
as emerging from the AL method~\cite{afonso2010fast}.
\begin{algorithm}[h]
    \caption{Fast Equivariant Imaging (FEI-Option 2)}
    \label{alg:FEI-O2}
    \begin{algorithmic}[1]
        \REQUIRE 
        \begin{tabular}[t]{@{}p{0.45\linewidth}@{}p{0.45\linewidth}@{}}
            $\bullet$ $\mathcal{Y}$ - Observations (training dataset)  & $\bullet$ $N$ - number of epochs \\
            $\bullet$ $\mathbf{L}$ - Dual variable   & $\bullet$ $\mathcal{G}$ - Group of transformations \\
            $\bullet$ $\lambda$ - Penalty parameter  & $\bullet$ $\mathbf{u}_0$ - Initial gradient descent point \\
            $\bullet$ $\alpha$ - EI regularization parameter           & $\bullet$ $\gamma$ - Gradient descent step size
        \end{tabular}
        
        \STATE \textbf{Initialize:} Neural network $\Ft$, $\mathbf{L}_0 = \mathbf{0}$, $\mathbf{u}_0 = \mathbf{0}$, $k=0$;
        
        \FOR{$i=0, \dots, N-1$}
            \FOR{$\mathbf{y} \in \mathcal{Y}$}
                \STATE Sample $\mathrm{T}_g$ where $g \sim \mathcal{G}$;
                \STATE Compute $\mathbf{x}_0 = \mathcal{F}_{\theta_k}(\mathbf{y})$, by default let $\mathbf{u}_k = \mathbf{x}_0$;
                \STATE Update $\mathbf{u}_{k+1} = \mathbf{u}_k - \gamma[  \nabla f_\mr{mc}( \mathbf{y} , \mathbf{A}\mathbf{u}_k) + \lambda (\mathbf{u}_k - \mathcal{F}_{\theta_k}(\mathbf{y}) + \mathbf{L}_k^{(\mathbf{y})}) ]$;
                \STATE Let $\mathbf{x}_1 = \mathbf{u}_{k+1}$;
                \STATE Transform $\mathbf{x}_2 = \mathrm{T}_g(\mathbf{x}_1)$;
                \STATE Compute $\mathbf{x}_3 = \mathcal{F}_{\theta}( \mathbf{A} \mathbf{x}_2)$;
                \STATE Update $\theta_{k+1}$ via progressively minimizing $\mathcal{L} = \mathbb{E}_{ g \sim \mathcal{G}} \{ \Vert \mathcal{F}_{\theta}(\mathbf{y}) - \mathbf{L}_k^{(\mathbf{y})} - \mathbf{x}_1 \Vert^2_2 + \alpha \Vert \mathbf{x}_2 - \mathbf{x}_3 \Vert^2_2 \}$  using an adaptive gradient-based optimizer (such as Adam) in each iteration;
                \STATE Update $\mathbf{L}_{k+1}^{(\mathbf{y})} = \mathbf{L}_k^{(\mathbf{y})} +  \mathbf{x}_1 - \mathcal{F}_{\theta_{k+1}}(\mathbf{y})$;
                \STATE $k \leftarrow k+1$
            \ENDFOR
        \ENDFOR
        
        \ENSURE $\mathcal{F}_{\theta^\star}$
    \end{algorithmic}
\end{algorithm}

\subsection{PnP-FEI: Utilizing Both Primal and Dual Priors}
FEI provides us with a powerful way of effectively and efficiently utilizing the dual-domain prior such as EI for unsupervised training of deep imaging networks. Since in FEI we split the training problem with a latent-reconstruction, we can also choose to utilize the primal (image) domain prior to further accelerate the training and improve the performance. The plug-and-play~\cite{venkatakrishnan2013plug} prior is a flexible framework that incorporates state-of-the-art denoising algorithms as priors in more challenging inverse problems. We extend this here to Fast Equivariant Imaging as an auxiliary primal regularization on the image domain, in the hope that joint-forcing with dual-domain EI regularization will further improve optimization speed and generalization performance in practice. 

An alternative approach to update $\mathbf{u}$ in \textbf{Latent-reconstruction} step of proposed FEI framework, is to include plug-and-play denoisers as external denoising prior:
\begin{equation}
    M_L(\mathbf{y}, \mathcal{F}_{\theta_k}) \leftarrow \mathcal{D}_{\sigma}(\mathbf{u}_k),
\end{equation}
where $\mathbf{u}_{k}$ is obtained by one step gradient descent of latent-reconstruction objective (\textit{e.g.} \eqref{eq: sub1}). and $\mathcal{D}(\cdot)$ represents classical denoisings such as BM3D~\cite{dabov2007image}, DnCNN~\cite{zhang2017beyond}, etc.
\begin{eqnarray*}
 && \mathrm{\textbf{PnP-FEI generic framework}}\\&& - \mathrm{Initialize}\ \theta_0\\
 &&\mathrm{For} \ \ \ k = 0, 1, 2,...,  K\\
&&\left\lfloor
\begin{array}{l}
\mathrm{Sample\ } \mathbf{y} \in \mathcal{Y}, g \sim \mathcal{G} \\
\mathbf{q}_{k} \leftarrow M_L(\mathbf{y}, \mathcal{F}_{\theta_k}) \\
\mathbf{u}_{k+1} \leftarrow \mathcal{D}_{\sigma}(\mathbf{q}_k)  \qquad \mathrm{\it\color{purple} PnP\ latent\ step}\\
\theta_{k+1} \leftarrow M_P(\mathbf{u}_{k+1}, \mathcal{F}_{\theta_k})  \ \  \mathrm{\it\color{purple} Pseudo\ supervision}
\end{array}
\right.
\end{eqnarray*}
The PnP-FEI framework summarizes the steps to be taken to apply this overall algorithm to manage the minimization of the objective of equivariant imaging. Here we present the PnP-FEI with ADMM-inspired relaxed splitting option (Option 2), while PnP-FEI Option 1 uses the same approach with denoising steps following the gradient steps in the latent-reconstruction (we omit writing this down for the sake of avoiding repeated description).

\begin{algorithm}[h]
    \caption{Fast Equivariant Imaging with Plug-and-Play (PnP-FEI Option 2)}
    \label{alg:PnP-FEI-O2}
    \begin{algorithmic}[1]
        \REQUIRE 
        \begin{tabular}[t]{@{}p{0.45\linewidth}@{}p{0.45\linewidth}@{}}
            $\bullet$ $\mathcal{Y}$ - Observations (training dataset)  & $\bullet$ $N$ - number of epochs \\
            $\bullet$ $\mathbf{L}$ - Dual variable  & $\bullet$ $\mathcal{G}$ - Group of transformations \\
            $\bullet$ $\lambda$ - Penalty parameter  & $\bullet$ $\mathbf{u}_0$ - Initial gradient descent point \\
            $\bullet$ $\alpha$ - EI regularization parameter           & $\bullet$ $\gamma$ - Gradient descent step size
        \end{tabular}
        
        \STATE \textbf{Initialize:} Neural network $\Ft$, $\mathbf{L}_0 = \mathbf{0}$, $\mathbf{u}_0 = \mathbf{0}$, $k=0$;
        
        \FOR{$i=0, \dots, N-1$}
            \FOR{$\mathbf{y} \in \mathcal{Y}$}
                \STATE Sample $\mathrm{T}_g$ where $g \sim \mathcal{G}$;
                \STATE Compute $\mathbf{x}_0 = \mathcal{F}_{\theta_k}(\mathbf{y})$, by default let $\mathbf{u}_k = \mathbf{x}_0$;
                \STATE Update $\mathbf{u}_{k+1} = {\color{black} \mathcal{D}_{\sigma}}[ \mathbf{u}_k - \gamma[  \nabla f_\mr{mc}( \mathbf{y} , \mathbf{A}\mathbf{u}_k) + \lambda (\mathbf{u}_k - \mathcal{F}_{\theta_k}(\mathbf{y}) + \mathbf{L}_k^{(\mathbf{y})}) ]]$;
                \STATE Let $\mathbf{x}_1 = \mathbf{u}_{k+1}$;
                \STATE Transform $\mathbf{x}_2 = \mathrm{T}_g(\mathbf{x}_1)$;
                \STATE Compute $\mathbf{x}_3 = \mathcal{F}_{\theta}( \mathbf{A} \mathbf{x}_2)$;
                \STATE Update $\theta_{k+1}$ via progressively minimizing $\mathcal{L} = \mathbb{E}_{ g \sim \mathcal{G}} \{ \Vert \mathcal{F}_{\theta}(\mathbf{y}) - \mathbf{L}_k^{(\mathbf{y})} - \mathbf{x}_1 \Vert^2_2 + \alpha \Vert \mathbf{x}_2 - \mathbf{x}_3 \Vert^2_2 \}$  using an adaptive gradient-based optimizer (such as Adam) in each iteration;
                \STATE Update $\mathbf{L}_{k+1}^{(\mathbf{y})} = \mathbf{L}_k^{(\mathbf{y})} +  \mathbf{x}_1 - \mathcal{F}_{\theta_{k+1}}(\mathbf{y})$;
                \STATE $k \leftarrow k+1$
            \ENDFOR
        \ENDFOR
        
        \ENSURE $\mathcal{F}_{\theta^\star}$
    \end{algorithmic}
\end{algorithm}

In this paper, PnP-FEI is treated purely as an empirical extension of FEI: it augments the latent-reconstruction step with an external image prior, but we do not claim a separate convergence result for the practical pretrained-denoiser update.

\subsection{Test Time Efficient Adaptation of FEI}
Equivariant Imaging (EI) is applicable not only to unsupervised learning but also to Test-Time Adaptation (TTA)~\cite{wang2024test}. Given the strict time constraints inherent in Test-Time Adaptation, the proposed FEI framework is a compelling candidate for accelerating the adaptation process, particularly in scenarios where EI is employed.

Let $\hat{\theta}$ denote the learned parameters of a pre-trained model. For a specific test instance $(\hat{\mathbf{y}}, \hat{\mathbf{A}})$, we initialize $\theta_0 = \hat{\theta}$ and iteratively fine-tune the model parameters using the alternating FEI scheme for $T$ iterations:
\begin{equation}
    \begin{cases}
        \mathbf{u} \leftarrow M_L(\hat{\mathbf{y}}, \mathcal{F}_{\theta})\\
        \theta \leftarrow M_P(\mathbf{u}, \mathcal{F}_{\theta})
    \end{cases}
\end{equation}
where $M_L$ and $M_P$ correspond to the Latent reconstruction (Eq.~\ref{eq: sub1}) and Pseudo supervision (Eq.~\ref{eq: sub2}) updates, respectively. Once done, the final adaptation is given by
\[
\hat{\mathbf{x}} = \mathcal{F}_{\theta}(\hat{\mathbf{y}})
\]
with the adapted parameters $\theta$. As we will see, our FEI framework is particularly effective in test-time model adaptation, ensuring improved imaging accuracy and robustness against distribution shifts.

\section{Numerical Experiments}\label{sec: num_exp}
We assess the effectiveness of the proposed framework using experiments conducted on
different inverse problems. The inverse problems we consider are sparse-view CT and inpainting. The configuration is as follows:

\noindent\textbf{(a) Metric.} To evaluate reconstruction quality, we report standard PSNR and SSIM metrics. Additionally, we adopt the equivariance metric proposed by~\cite{chaman2021truly} to quantify the equivariance properties of the estimator. This metric measures the mean squared error deviation from the canonical equivariance constraint $\mathrm{T}_g G_\theta(\mathbf{Ax}) = G_\theta(\mathbf{A} \mathrm{T}_g\mathbf{x})$, expressed in dB:
\begin{equation}
    \mathrm{EQUIV} = -10 \log \left( \mathbb{E}_{g} \left\{ \| \mathrm{T}_g \mathcal{F}_\theta(\mathbf{Ax}) -  \mathcal{F}_\theta(\mathbf{A}\mathrm{T}_g\mathbf{x}) \|_2^2 \right\} \right),
\end{equation}
Intuitively, this metric functions as a PSNR analogue specifically for quantifying equivariance.

\noindent\textbf{(b) Neural Network.} We use the same U-Net to bulid $G_{\theta}$ as suggested in~\cite{chen2021equivariant} for EI, FEI and PnP-FEI. Specifically, $G_{\theta}$ is a four-depth residual U-Net~\cite{ronneberger2015u} with a symmetric encoder-decoder and long skip connections. Each encoder comprises two $3 \times 3$ Convolution-BatchNorm-ReLU blocks with identity shortcuts, followed by $2 \times 2$ max pooling to double channels $[64, 128, 256, 512]$. The decoder mirrors the encoder in a refinement manner, and a final $1 \times 1$ convolution restores the output. For the measurement consistency loss, for simplicity, we choose $\ell_2$ in this experiment.

\noindent\textbf{(c) Baselines.} For the training phase, we benchmark our proposed FEI scheme against: (i) standard EI; and (ii) PnP-FEI instantiated with a pre-trained DnCNN denoiser (see Table~\ref{tab:vs-strategy}). We set the noise standard deviation to $\sigma = 0.01$ and utilize the implementations provided by the \texttt{DeepInv} library\footnote{\url{https://deepinv.github.io/deepinv/}}. To ensure reproducibility and fairness, all EI results were generated using the official source code released by~\cite{chen2021equivariant}. Regarding Test-Time Adaptation (TTA), we evaluate the proposed approach against three distinct strategies: (i) Standard Test-Time Training (TTT)~\cite{darestani2022test}, which relies solely on measurement consistency loss; (ii) A modified AdaptNet~\cite{zhao2024test}, where we integrate the original lightweight adaptive layer into our U-Net backbone, optimizing it via equivariant loss to maintain architectural consistency; and (iii) our proposed FEI adaptation strategy.

\begin{table*}[h!]
\centering
\caption{Unsupervised learning algorithms in our experiments.}
\label{tab:vs-strategy}
\begin{tabularx}{\linewidth}{ccX}
\toprule
\textbf{Framework} & \textbf{Learning strategy} & \textbf{Description} \\
\midrule
EI & Adam & Vanilla EI framework, which use back-propagation and end-to-end training strategy. \\
\addlinespace 
FEI-Option1 & Inexact HQS + Adam & HQS-inspired splitting with accelerated gradient descent to approximately minimize Latent-reconstruction step subproblem, while Adam for online Pseudo-supervision, Algorithm~\ref{alg:LEI}. \\
\addlinespace
    FEI-Option2 & Linearized ADMM + Adam & The difference from FEI-Option1 is the change of splitting style to linearized ADMM, with gradient step for Latent-reconstruction, while Adam for online Pseudo-supervision, Algorithm~\ref{alg:FEI-O2}. \\
\addlinespace
PnP-FEI & Dual-domain + Adam & The difference from FEI is the use of pre-trained denoiser for Latent-reconstruction step. \\
\bottomrule
\end{tabularx}
\end{table*}

\noindent\textbf{(d) Initialization.} For consistency, all learnable weights are initialized following the protocol of~\cite{chen2021equivariant}. In the proposed FEI framework for sparse-view CT, we configure the hyperparameters as follows: quadratic penalty parameter $\lambda=1$, NAG momentum $\beta=0.1$, step size $\eta=0.01$, and inner iterations $J=10$. Conversely, for the image inpainting task, we set $\lambda=0.1$, $\beta=0.9$, $\eta=0.09$, and maintain $J=10$.

\noindent\textbf{(e) Training.} For sparse-view CT, we utilize the Adam optimizer initialized with a learning rate of $1 \times 10^{-3}$, adhering to the decay schedule outlined in~\cite{chen2021equivariant}. We train the models for 5,000 epochs, employing a batch size of 8 and an equivariance strength $\alpha=1000$ to balance performance and efficiency. For the image inpainting task, we maintain the initial learning rate at $1 \times 10^{-3}$ with a total training duration of 2,000 epochs. We configure the batch size to 4 and set the equivariance strength to $\alpha=1.0$.  To enhance stability, we adopt the moving average technique from~\cite{sun2021plug} to smooth the reconstructed outputs. Finally, we implement early stopping with a patience of 300 epochs to mitigate overfitting.

\noindent\textbf{(f) Adaptation.} We evaluate the proposed approach on sparse-view CT images for TTA task, under two distinct conditions: minor distribution shifts and severe out-of-distribution (OOD) scenarios. We utilize a publicly available pre-trained model trained within the EI framework. Specifically, we employ the model from the original study by~\cite{chen2021equivariant}, which was trained on 90 clean samples from the CT-100 dataset for $T=5000$ epochs using the Adam optimizer with a constant learning rate of $5 \times 10^{-4}$. The adaptation process utilizes the Adam optimizer with a fixed learning rate of $1 \times 10^{-3}$ for $T = 100$ epochs following the procedure in~\cite{zhao2024test, quan2022dual}.

\subsection{Sparse-view CT}
The forward operator $\mathbf{A}$ for sparse-view CT is modeled as the Radon transform , where 50 projection angles are uniformly subsampled to generate sparse-view sinograms (observations) $\mathbf{y}$. The transformation group $G$ employed for equivariance consists of 2D planar rotations, implemented by sampling random angles from the full $360^{\circ}$ range. For the dataset, we utilize 90 images sampled from the CT100~\cite{clark2013cancer} dataset for training. The remaining 10 images, processed with the same scanning geometry, are reserved to evaluate the generalization performance of the trained models. All images are resized to a resolution of $128 \times 128$ (16384) pixels.

\begin{figure}[h!]
    \centering
    \includegraphics[width=1.0\linewidth]{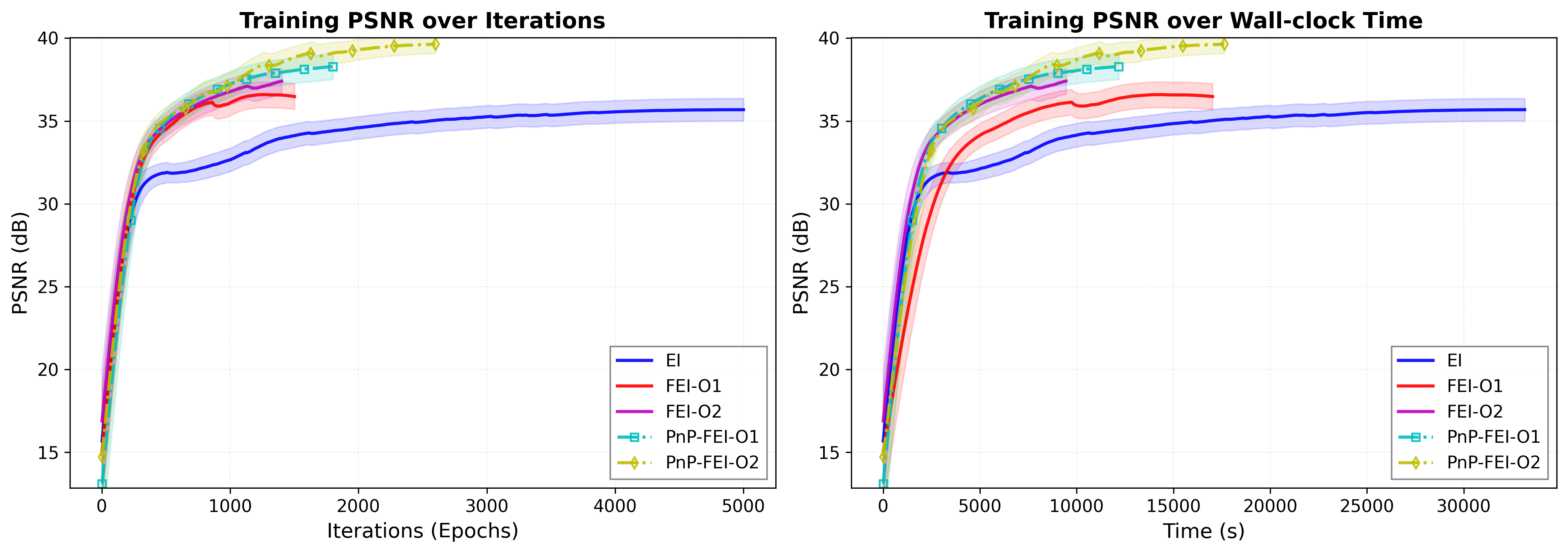}
    \includegraphics[width=1.0\linewidth]{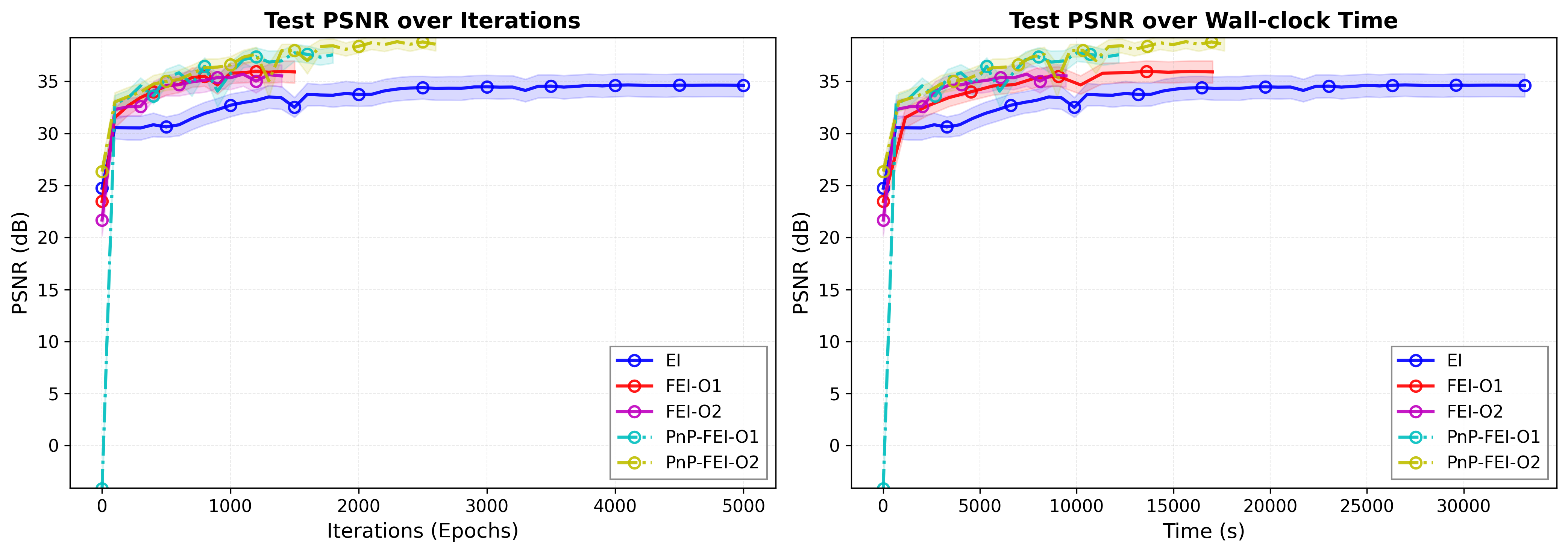}
    \caption{PSNR and MSE curves with respect to iteration and time for Sparse-view CT reconstruction. We observe a 10-time acceleration can be achieved by FEI over EI.}
    \label{fig:ct-psnr&mse-iters5000-curve}
\end{figure}

\begin{figure}[h!]
    \centering
    \includegraphics[width=1.0\linewidth]{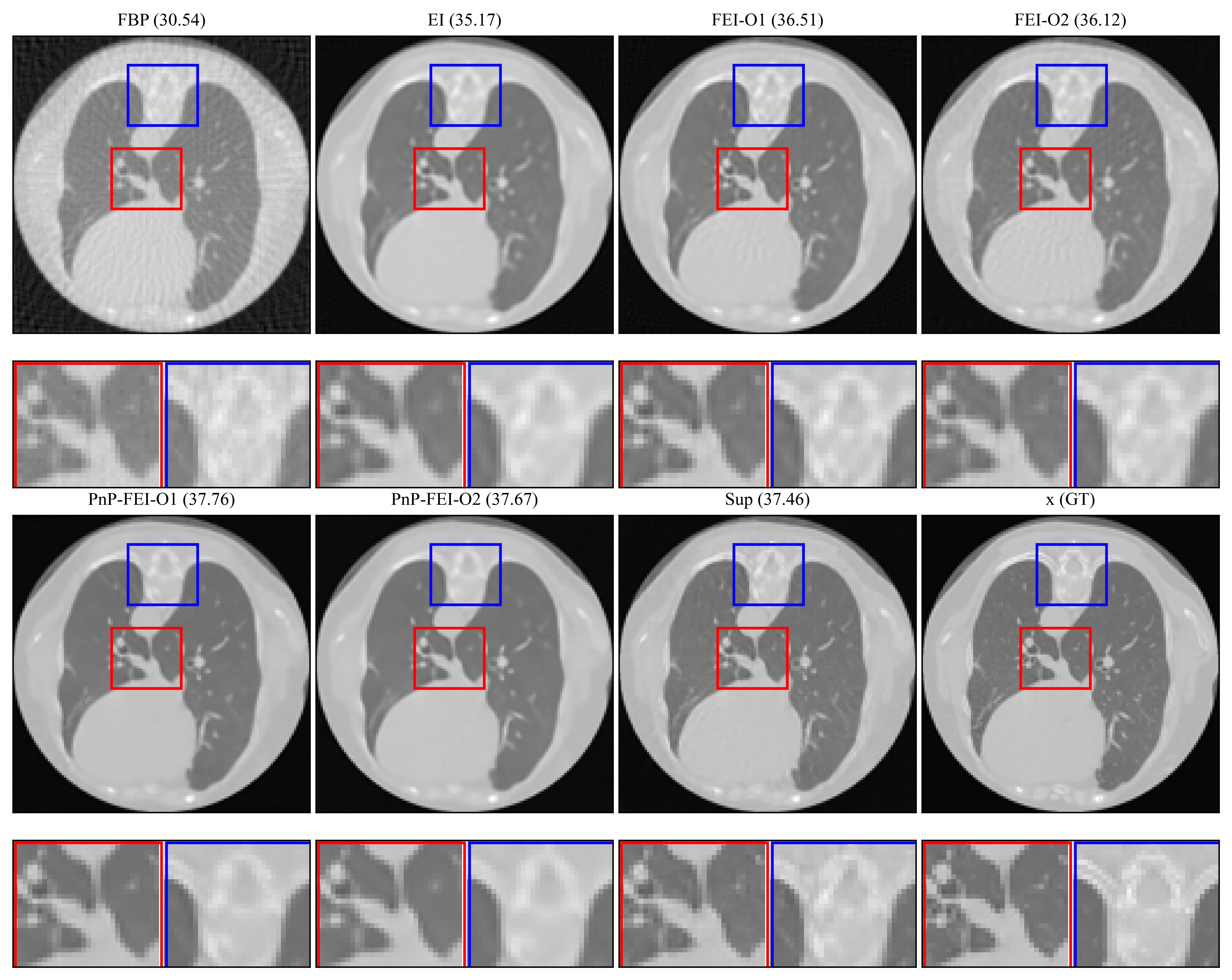}
    \caption{Visualization of test sample for sparse-view CT reconstruction.}
    \label{fig:50-ct-test}
\end{figure}

To evaluate the efficiency gains of the proposed FEI scheme for sparse-view CT imaging, we show the PSNR curves with respect to both iterations and times in Figure~\ref{fig:ct-psnr&mse-iters5000-curve}. Notably, FEI achieves an order-of-magnitude ($\geq10\times$) acceleration over standard EI training. { Despite the inexact nature of the latent variable update, FEI exhibits a smooth and stable training trajectory similar to the exact EI baseline. This empirically suggests that the approximation error introduced by our relaxed splitting does not destabilize the optimization process.} Furthermore, Table~\ref{tab:ct-results} and Figure~\ref{fig:50-ct-test} show that networks trained via our FEI and PnP-FEI strategies exhibit superior generalization performance compared to standard EI baselines.

\begin{table}[h!]
  \centering
  \small
  \setlength{\tabcolsep}{10pt}
  \caption{Averaged PSNR and SSIM on testing dataset for sparse-view CT reconstruction.}
  \begin{tabular}{lcc}
    \toprule
    Method & PSNR $\uparrow$ & SSIM $\uparrow$ \\
    \midrule
    Supervised & $38.17 \pm 1.57$ & $0.9689 \pm 0.0073$ \\
    \midrule
    PnP-FEI-O2 & $\mathbf{37.56 \pm 1.33}$ & $0.9648 \pm 0.0053$ \\
    PnP-FEI-O1 & $37.35 \pm 1.43$ & $\mathbf{0.9666 \pm 0.0054}$ \\
    FEI-O1 & $36.17 \pm 1.44$ & $0.9579 \pm 0.0065$ \\
    FEI-O2 & $35.94 \pm 1.65$ & $0.9528 \pm 0.0093$ \\
    EI & $35.03 \pm 1.53$ & $0.9531 \pm 0.0067$  \\
    \midrule
    FBP & $30.24 \pm 1.33$ & $0.8335 \pm 0.0341$  \\
    \bottomrule
  \end{tabular}
  \label{tab:ct-results}
\end{table}

\subsection{Image Inpainting}
\begin{figure}[h!]
    \centering
    \includegraphics[width=1.0\linewidth]{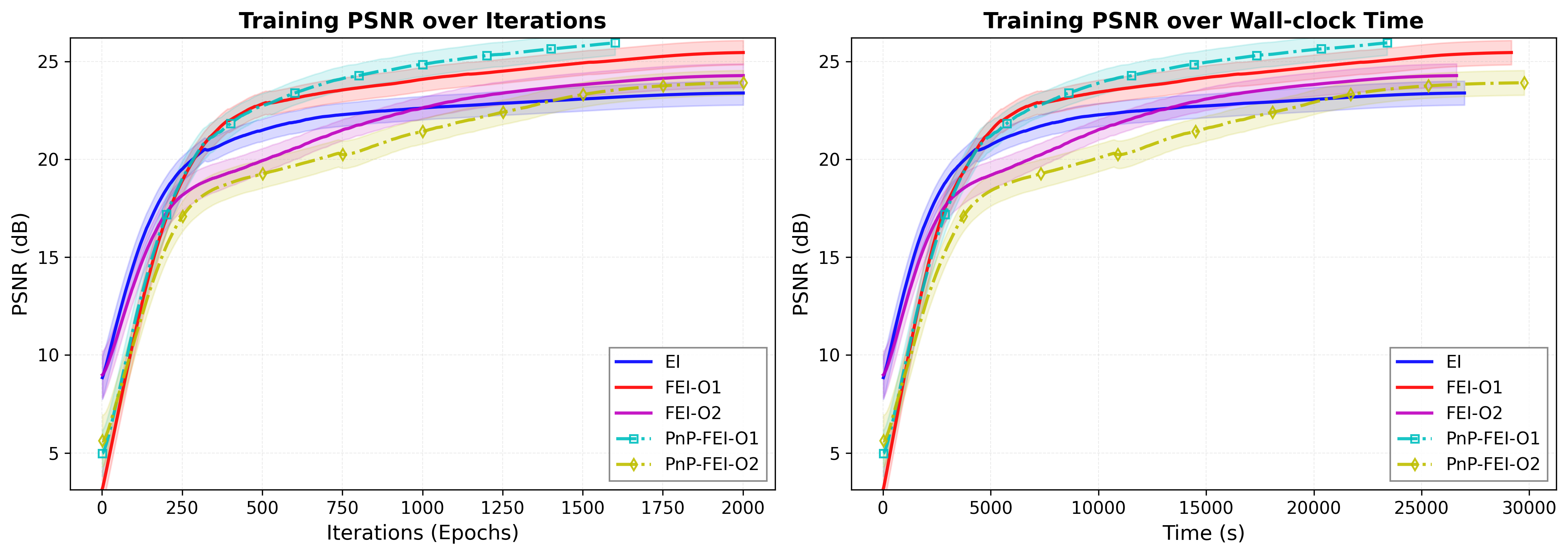}
    \includegraphics[width=1.0\linewidth]{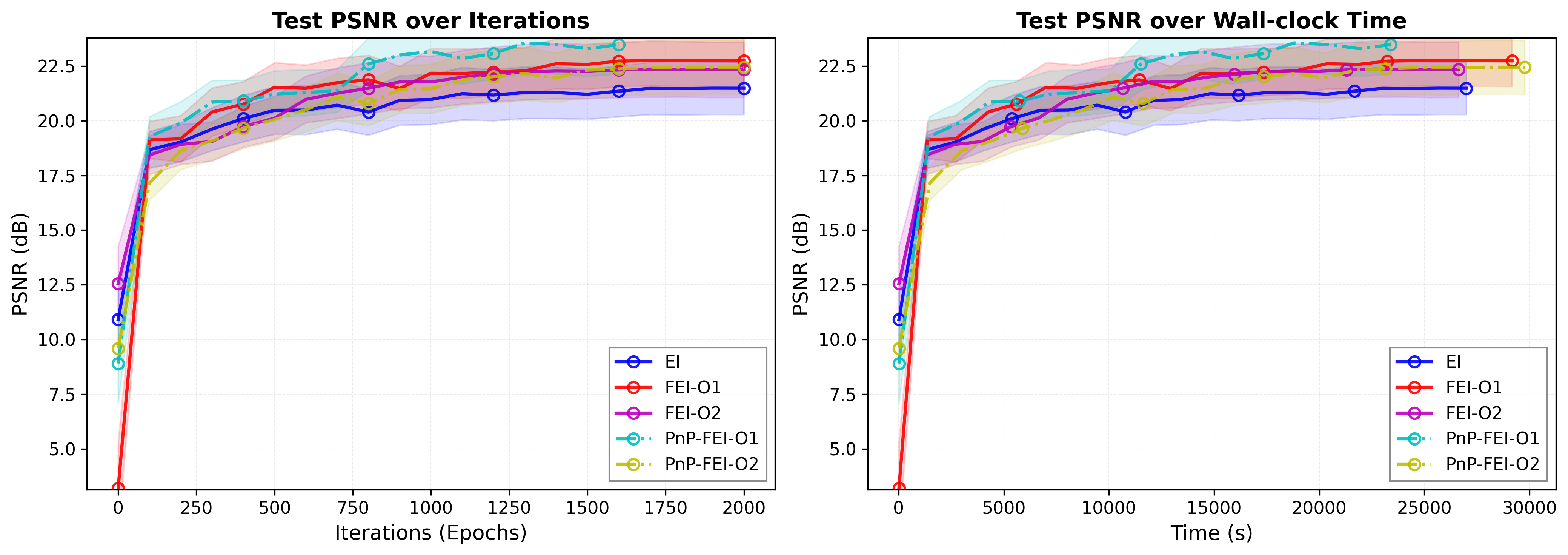}
    \caption{PSNR and MSE curves with respect to iteration and time for image inpainting. We observe an significant acceleration can be achieved by FEI over EI.}
    \label{fig:inpainting-psnr&mse-iters2000-curve}
\end{figure}

\begin{figure}[h!]
    \centering
    \includegraphics[width=1.0\linewidth]{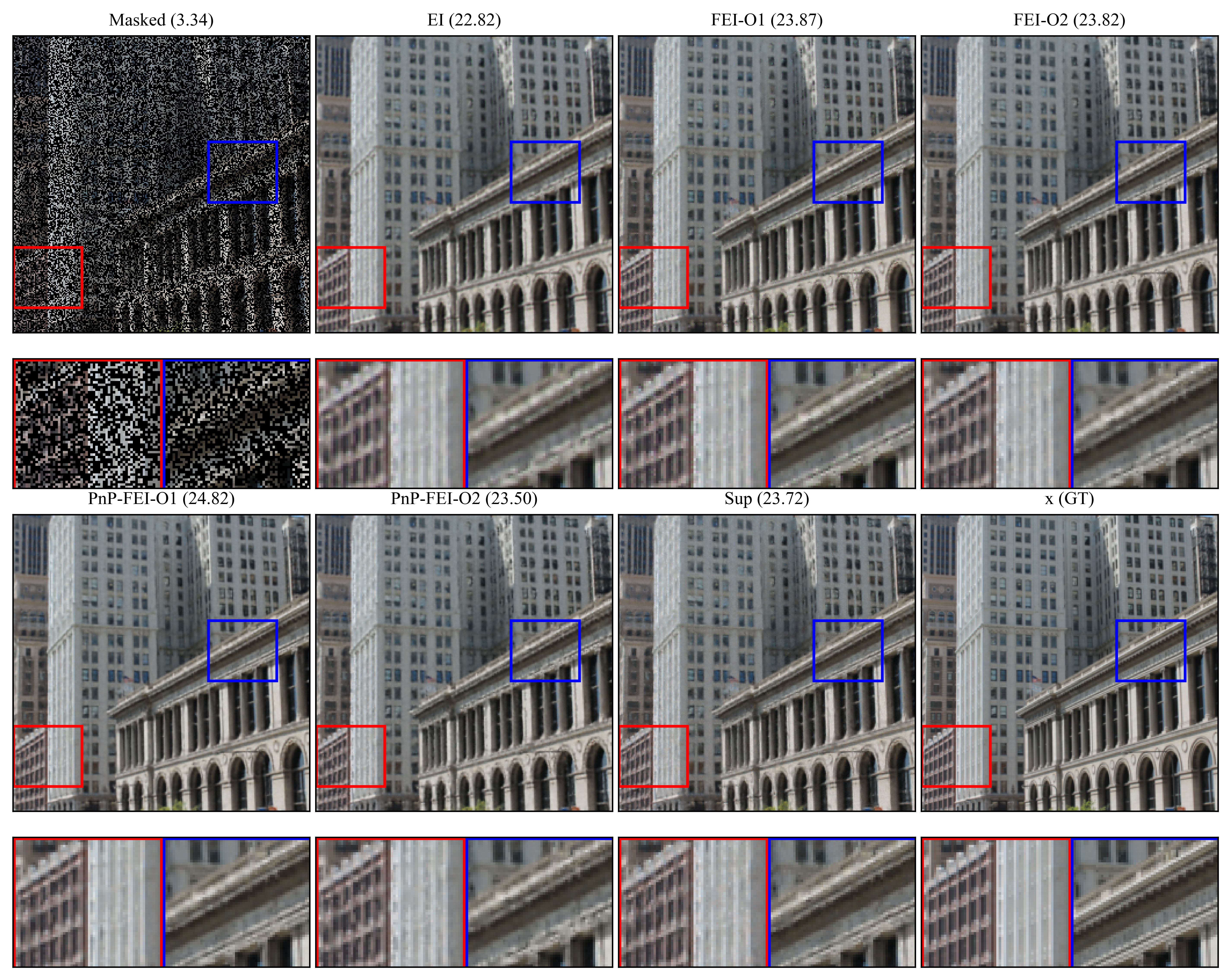}
    \caption{Visualization of test sample for imaging inpainting.}
    \label{fig:inpaint-test}
\end{figure}

The mask rate for the inpainting task is $0.6$, which means 60\% of pixels are randomly removed. Here we choose random shift as the transformation. The dataset we used here is Urban100~\cite{huang2015single} which is the same as~\cite{chen2021equivariant}. For each image, we cropped a $512 \times 512$ pixel area in the center and then resized it to $256 \times 256$ ($65536$ pixels) for the ground truth image. The first 90 measurements are for training, while the last 10 images are for testing. Figure~\ref{fig:inpainting-psnr&mse-iters2000-curve} demonstrates significantly faster improvement compared to the vanilla EI method. Furthermore, Table~\ref{tab:inpainting-results} and Figure~\ref{fig:inpaint-test} indicate that the FEI-based approach achieves superior generalization performance relative to the standard EI method.

\begin{table}[h!]
  \centering
  \caption{Averaged PSNR and SSIM on testing dataset for imaging inpainting.}
  \small
  \setlength{\tabcolsep}{10pt}
  \begin{tabular}{lcc}
    \toprule
    Method & PSNR $\uparrow$ & SSIM $\uparrow$ \\
    \midrule
    Supervised & $22.65 \pm 2.15$ & $0.8928 \pm 0.0434$ \\
    \midrule
    PnP-FEI-O1 & $\mathbf{23.56 \pm 2.56}$ & $\mathbf{0.9116 \pm 0.0482}$ \\
    PnP-FEI-O2 & $22.25 \pm 2.41$ & $0.8913 \pm 0.0559$ \\
    FEI-O1 & $22.75 \pm 2.22$ & $0.9043 \pm 0.0500$ \\
    FEI-O2 & $22.64 \pm 2.38$ & $0.8971 \pm 0.0522$ \\
    EI & $21.49 \pm 2.30$ & $0.8733 \pm 0.0650$  \\
    \midrule
    Masked & $2.84 \pm 1.57$ & $0.1196 \pm 0.0241$  \\
    \bottomrule
  \end{tabular}
  \label{tab:inpainting-results}
\end{table}

\subsection{Additional Analysis on Test-Time Model Adaptation}
In this section, we evaluate the proposed FEI framework for TTA with sparse-view CT reconstruction task. Considering the practical applications, in the following experiments, we assume that the noise distribution is mixed Poisson-Gaussian, i.e.,
 \begin{equation}
     \mathbf{y} = \gamma \mathcal{P}(\frac{\mathbf{x}}{\gamma}) + \sigma \epsilon,
 \end{equation}
 where $\mathcal{P}$ denotes the Poisson distribution, and $\epsilon \in \mathcal{N}(\mathbf{0}, \mathbf{I})$.

\begin{figure}[h]
  \centering
  \begin{minipage}[b]{0.48\linewidth}
    \centering
    \includegraphics[width=\linewidth]{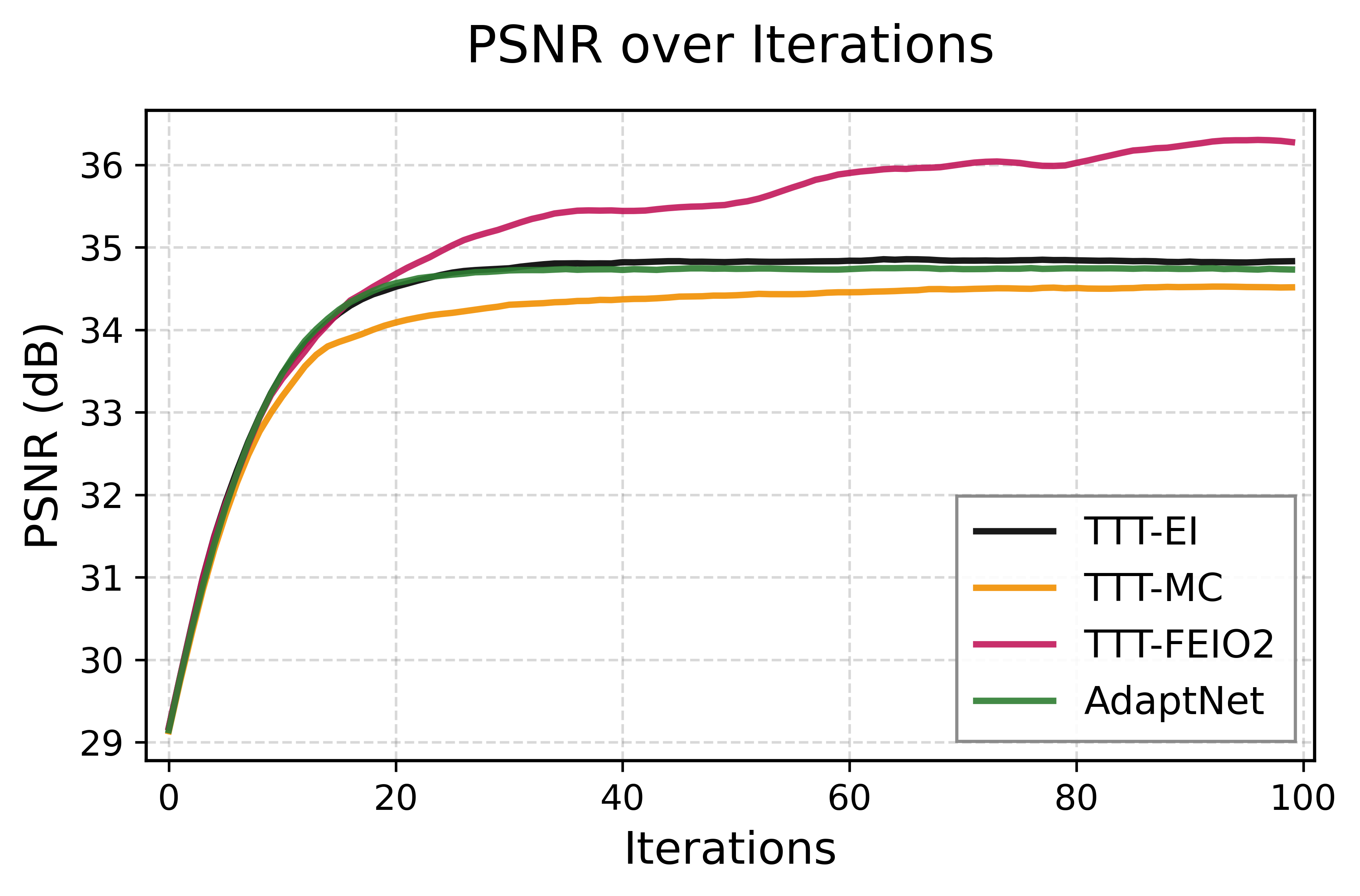} 
    \vspace{4pt}
    {\footnotesize (a) Mean PSNR values vs. training epochs.}
  \end{minipage}\hfill
  \begin{minipage}[b]{0.48\linewidth}
    \centering
    \includegraphics[width=\linewidth]{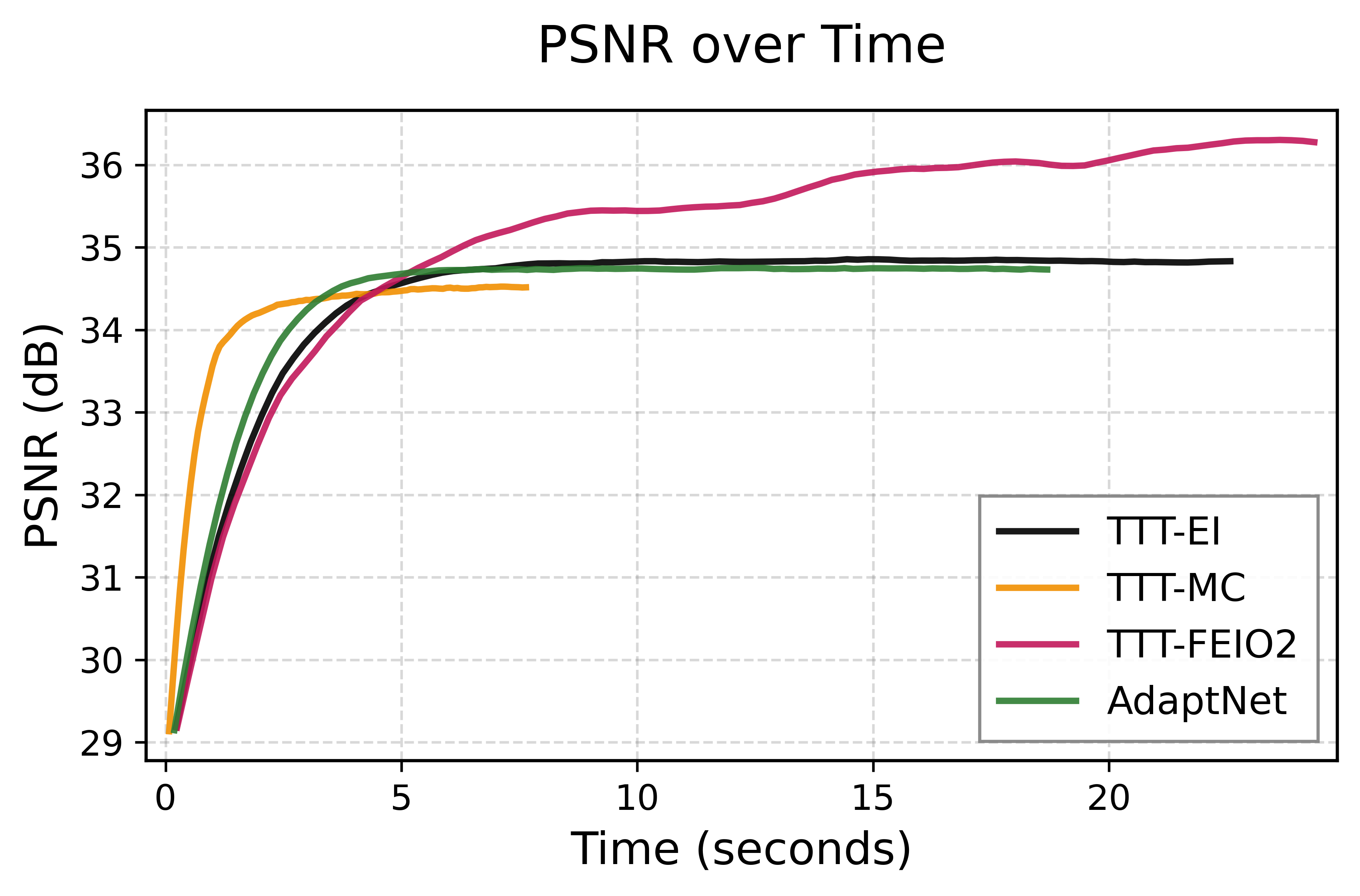} 
    \vspace{4pt}
    {\footnotesize (b) Mean PSNR values vs. wall clock time.}
  \end{minipage}
  \caption{PSNR curves vs. iteration numbers and time during adaptation for different methods.}
  \label{fig:fei-ttt-comp}
\end{figure}

To evaluate the efficacy of the proposed method in Test-Time Adaptation (TTA), we conducted comparative experiments across two distinct scenarios. First, to assess robustness against mild distribution shifts, we benchmarked various adaptation methods under varying noise intensities. As shown in Table~\ref{tab: fei-ttt-iid} and Figure~\ref{fig:fei-ttt-comp}, FEI demonstrates superior adaptation capabilities under minor shifts; it achieves state-of-the-art reconstruction performance while maintaining competitive computational efficiency. Even as noise intensity escalates, FEI consistently delivers high-fidelity reconstructions, underscoring its robustness in low-shift regimes. 

\begin{figure*}[h!]
    \centering
    \includegraphics[width=1.0\linewidth]{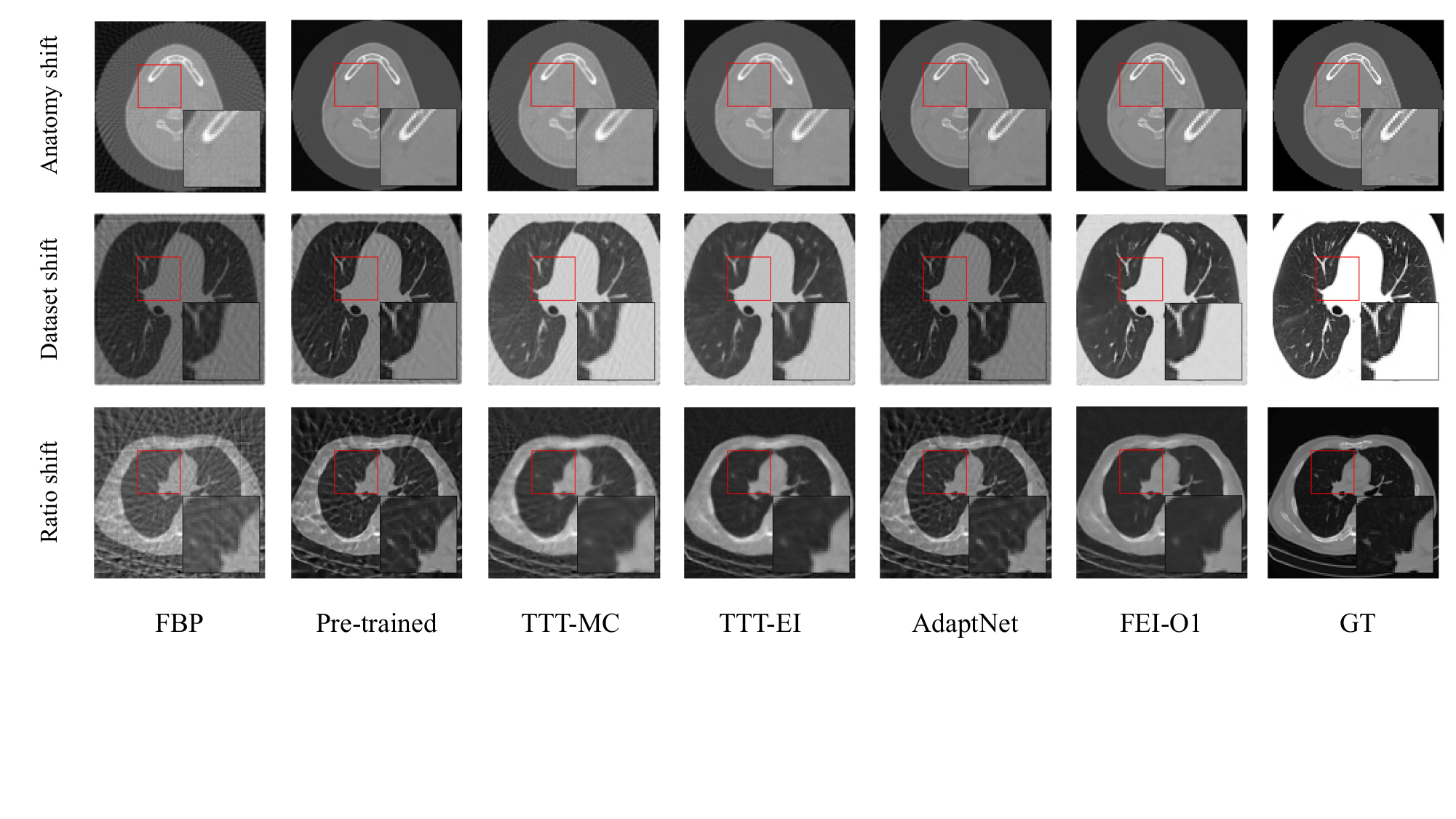}
    \caption{Visualization results of different adaptation approaches across three domain shift scenarios.}
    \label{fig:fei-ttt}
\end{figure*}

Subsequently, to rigorously evaluate FEI under significant distribution shifts, we extended our comparison to scenarios involving substantial domain discrepancies. Specifically, we defined the source domain $\mathcal{S}$ and target domain $\mathcal{D}$ for three specific shifts: 1) \textit{Anatomy shift}, where $\mathcal{S}$ comprises body (lung and abdomen) CT images collected from Mayo at 50 views, and $\mathcal{D}$ consists of brain CT images from the same setting; 2) \textit{Dataset shift}, where $\mathcal{S}$ denotes body images from Mayo, and $\mathcal{D}$ represents the SARS-COV2 dataset; and 3) \textit{Ratio shift}, where $\mathcal{S}$ contains body images from Mayo at 50 views, while $\mathcal{D}$ contains body images from Mayo at 25 views. As can be seen from Table~\ref{tab:fei-ttt-ood} and Figure~\ref{fig:fei-ttt}, FEI achieves superior performance against other methods across all three domain shift scenarios.

\begin{table*}[t]
\centering
\caption{Mean PSNR values of sparse-view CT reconstructions with different noise levels for TTA. Pre-trained model here is the pre-trained model with the same backbone as EI. The best performance is highlighted in \textbf{bold}.} 
\resizebox{\textwidth}{!}{%
\begin{tabular}{|l|c|c|ccc|c|ccc|}
\hline
\hline
\multirow{4}{*}{Noise} & \multirow{4}{*}{$\sigma, \gamma$} & \multicolumn{1}{c|}{Classic} & \multicolumn{3}{c|}{Unsupervised} & TTT* & \multicolumn{3}{c|}{Network Adaptation} \\ \cline{3-10} 
 &  & FBP & Pre-trained & \makecell[c]{REI \\ (known $\sigma, \gamma$) \\ \cite{chen2022robust}} & \makecell[c]{EI-UNSURE \\(unknown $\sigma, \gamma$) \\ \cite{tachella2024unsure}} & \makecell[c]{MC \\ (unknown $\sigma, \gamma$) \\ \cite{darestani2022test}} & \makecell[c]{EI \\(unknown $\sigma, \gamma$) \\ \cite{chen2021equivariant}} & \makecell[c]{AdaptNet \\(unknown $\sigma, \gamma$) \\ \cite{zhao2024test}} & \makecell[c]{FEI-O2 \\ (unknown $\sigma, \gamma$)}  \\ 
 \hline
\multirow{3}{*}{MPG} & $5 \times 10^{-3}$ & 25.55 $\pm$ 0.83 & 30.26 $\pm$ 1.19 & 32.93 $\pm$ 0.69 & 30.88 $\pm$ 0.70 & 34.47 $\pm$ 1.02 & 34.71 $\pm$ 0.95 & 34.74 $\pm$ 1.12 & \textbf{36.04 $\pm$ 1.41} \\
 & $1 \times 10^{-2}$ & 23.49 $\pm$ 1.13 & 28.10 $\pm$ 1.43 & 30.78 $\pm$ 0.66 & 28.11 $\pm$ 0.82  & 33.65 $\pm$ 0.88 & 33.70 $\pm$ 0.91 & 33.50 $\pm$ 1.06  & \textbf{34.95 $\pm$ 0.88}  \\
  & $5 \times 10^{-2}$ &  17.44 $\pm$ 1.57 & 22.02 $\pm$ 1.86 & 28.46 $\pm$ 0.75  & 22.42 $\pm$ 0.92 & 30.35 $\pm$ 0.80 &  \textbf{31.30 $\pm$ 0.82} & 29.12 $\pm$ 1.15 & 31.26 $\pm$ 0.78  \\
 \hline
\end{tabular}}
\label{tab: fei-ttt-iid}
\end{table*}

\begin{table*}[t]
\centering
\caption{Mean PSNR values of sparse-view CT across three domain shift scenarios with different adaptation approaches. The best is highlighted in \textbf{bold}.}
\setlength{\tabcolsep}{4pt}
\resizebox{0.75\textwidth}{!}{%
\begin{tabular}{lccccc}
\toprule
\textbf{Algorithm} & \textbf{Setup} & \textbf{Methods} & \textbf{Anatomy Shift} & \textbf{Dataset Shift} & \textbf{Ratio Shift} \\
& & & $\mathcal{S}$: Body & $\mathcal{S}$: Mayo & $\mathcal{S}$: 50 views \\
& & &$\mathcal{D}$: Brain & $\mathcal{D}$: SARS & $\mathcal{D}$: 25 views \\
\midrule
FBP & - & - & 30.01 $\pm$ 1.12  & 20.41 $\pm$ 0.87  & 24.78 $\pm$ 1.45  \\
\midrule
Pre-trained & \begin{tabular}[c]{@{}l@{}}Train on $\mathcal{S}$\\ Test on $\mathcal{D}$\end{tabular} & UNet & 35.71 $\pm$ 1.45  & 24.85 $\pm$ 1.34  & 27.02 $\pm$ 1.49
\\
\midrule
MC & \begin{tabular}[c]{@{}l@{}}Train on $\mathcal{S}$\\ TTA on $\mathcal{D}$\end{tabular} & UNet & 35.12 $\pm$ 1.77 & 31.14 $\pm$ 2.04  & 30.45 $\pm$ 1.76
\\
\midrule
EI & \begin{tabular}[c]{@{}l@{}}Train on $\mathcal{S}$\\ TTA on $\mathcal{D}$\end{tabular} & UNet & 35.98 $\pm$ 1.61 & 32.84 $\pm$ 2.20  & 32.02 $\pm$ 1.41 \\
\midrule
EI & \begin{tabular}[c]{@{}l@{}}Train on $\mathcal{S}$\\ TTA on $\mathcal{D}$ \end{tabular} & AdaptNet & 35.54 $\pm$ 1.36 & 23.78 $\pm$ 1.34 & 28.00 $\pm$ 1.61  \\
\midrule
FEI-O1 & \begin{tabular}[c]{@{}l@{}}Train on $\mathcal{S}$\\ TTA on $\mathcal{D}$ \end{tabular} & UNet & \textbf{37.82 $\pm$ 1.23} & \textbf{33.42 $\pm$ 1.74} & \textbf{32.23 $\pm$ 1.50}  \\
\bottomrule
\end{tabular}}
\label{tab:fei-ttt-ood}  
\end{table*}

\section{Conclusion}

We propose FEI and its variant PnP-FEI for accelerating unsupervised training of deep neural networks to imaging inverse problems without the need of ground-truth data. We observe that our framework is capable of significantly accelerating the state-of-the-art unsupervised training scheme Equivariant Imaging proposed by \cite{chen2021equivariant}, along with improved generalization performance and boosted adaptation performance. We believe that this work is crucial for the area of deep learning for computational imaging, via making the unsupervised training truly practical. Our algorithmic framework (inexact splitting plus PnP) can be easily extended to more recent variants of EI such as Robust Equivariant Imaging (REI) \cite{chen2022robust}, Multi-Operator Imaging (MOI) \cite{tachella2022unsupervised}, and our Sketched Equivariant Imaging (SkEI) \cite{xu2024sketched}, etc.

\appendices

\section{Proof of Proposition~\ref{lem:main_omitted_gradient_residual}}
\label{app:omitted_gradient_proof}

This appendix proves the residual-control proposition from Section~\ref{subsec:fei_theory}.

\begin{proof}[Proof of Proposition~\ref{lem:main_omitted_gradient_residual}]
Let
\[
R_{\rm eq}(\mathbf{u},\theta)
=
\mathrm{T}_g \mathbf{u}-\Ft(\mathbf{A}\mathrm{T}_g \mathbf{u}).
\]
Then
\[
\Psi(\mathbf{u},\theta)=\alpha\|R_{\rm eq}(\mathbf{u},\theta)\|_2^2,
\]
and
\[
D_{\mathbf{u}}R_{\rm eq}(\mathbf{u},\theta)
=
\mathrm{T}_g
-
J_{\mathbf{x}}\Ft(\mathbf{A}\mathrm{T}_g \mathbf{u})\mathbf{A}\mathrm{T}_g .
\]
Hence
\[
\nabla_{\mathbf{u}}\Psi(\mathbf{u},\theta)
=
2\alpha
\left[
\mathrm{T}_g
-
J_{\mathbf{x}}\Ft(\mathbf{A}\mathrm{T}_g \mathbf{u})\mathbf{A}\mathrm{T}_g
\right]^*
R_{\rm eq}(\mathbf{u},\theta).
\]
Taking norms and using \(\|J_{\mathbf{x}}\Ft(\mathbf{A}\mathrm{T}_g \mathbf{u})\|\leq C_x\), we obtain
\begin{align}
\|\nabla_{\mathbf{u}}\Psi(\mathbf{u},\theta)\|_2
&\leq
2\alpha
\left(
\|\mathrm{T}_g\|
+
\|J_{\mathbf{x}}\Ft(\mathbf{A}\mathrm{T}_g \mathbf{u})\|\,\|\mathbf{A}\|\,\|\mathrm{T}_g\|
\right)
\nonumber\\
&\qquad\qquad \times
\|R_{\rm eq}(\mathbf{u},\theta)\|_2
\nonumber\\
&\leq
2\alpha \|\mathrm{T}_g\|
\left(
1+\|\mathbf{A}\|C_x
\right)\|R_{\rm eq}(\mathbf{u},\theta)\|_2 .
\end{align}
This proves \eqref{eq:e_residual_bound_general}.
If \(\mathrm{T}_g\) is an isometry, then \(\|\mathrm{T}_g\|=1\), giving \eqref{eq:e_residual_bound_isometry}.
\end{proof}

\bibliographystyle{IEEEtran}
\bibliography{references}

\end{document}

%% file: fei_shared.tex
\ifCLASSOPTIONcompsoc
  \usepackage[nocompress]{cite}
\else
  \usepackage{cite}
\fi

\usepackage{amsmath,amssymb,amsfonts,amsthm,amsopn}
\usepackage{graphicx}
\usepackage{algorithm}
\usepackage{algorithmic}
\usepackage{multicol}
\usepackage{multirow}
\usepackage{tabularx}
\usepackage{booktabs}
\usepackage{makecell}
\usepackage{array}
\usepackage{xcolor}
\usepackage{url}
\usepackage{cleveref}

\ifCLASSINFOpdf
  \DeclareGraphicsExtensions{.pdf,.png,.jpg}
\else
  \DeclareGraphicsExtensions{.eps}
\fi

\theoremstyle{plain}

\newtheorem{proposition}{Proposition}

\theoremstyle{remark}

\crefname{hypothesis}{Hypothesis}{Hypotheses}
\crefname{fact}{Fact}{Facts}

\newcommand{\BEAS}{\begin{eqnarray*}}
\newcommand{\EEAS}{\end{eqnarray*}}
\newcommand{\BEA}{\begin{eqnarray}}
\newcommand{\EEA}{\end{eqnarray}}

\newcommand{\BEQ}{\begin{equation}}
\newcommand{\EEQ}{\end{equation}}
\newcommand{\BIT}{\begin{itemize}}
\newcommand{\EIT}{\end{itemize}}
\newcommand{\BNUM}{\begin{enumerate}}
\newcommand{\ENUM}{\end{enumerate}}

\newcommand{\Ft}{\mathcal{F}_\theta}

\newcommand{\mr}{\mathrm}

\newcommand{\BA}{\begin{array}}
\newcommand{\EA}{\end{array}}

%% file: main.bbl
\begin{thebibliography}{10}
\providecommand{\url}[1]{#1}
\csname url@samestyle\endcsname
\providecommand{\newblock}{\relax}
\providecommand{\bibinfo}[2]{#2}
\providecommand{\BIBentrySTDinterwordspacing}{\spaceskip=0pt\relax}
\providecommand{\BIBentryALTinterwordstretchfactor}{4}
\providecommand{\BIBentryALTinterwordspacing}{\spaceskip=\fontdimen2\font plus
\BIBentryALTinterwordstretchfactor\fontdimen3\font minus \fontdimen4\font\relax}
\providecommand{\BIBforeignlanguage}[2]{{%
\expandafter\ifx\csname l@#1\endcsname\relax
\typeout{** WARNING: IEEEtran.bst: No hyphenation pattern has been}%
\typeout{** loaded for the language `#1'. Using the pattern for}%
\typeout{** the default language instead.}%
\else
\language=\csname l@#1\endcsname
\fi
#2}}
\providecommand{\BIBdecl}{\relax}
\BIBdecl

\bibitem{xin2016maximal}
B.~Xin, Y.~Wang, W.~Gao, D.~Wipf, and B.~Wang, ``Maximal sparsity with deep networks?'' \emph{Advances in Neural Information Processing Systems}, vol.~29, 2016.

\bibitem{zhang2018ista}
J.~Zhang and B.~Ghanem, ``Ista-net: Interpretable optimization-inspired deep network for image compressive sensing,'' in \emph{Proceedings of the IEEE conference on computer vision and pattern recognition}, 2018, pp. 1828--1837.

\bibitem{belthangady2019applications}
C.~Belthangady and L.~A. Royer, ``Applications, promises, and pitfalls of deep learning for fluorescence image reconstruction,'' \emph{Nature methods}, vol.~16, no.~12, pp. 1215--1225, 2019.

\bibitem{ulyanov2018deep}
D.~Ulyanov, A.~Vedaldi, and V.~Lempitsky, ``Deep image prior,'' in \emph{Proceedings of the IEEE Conference on Computer Vision and Pattern Recognition}, 2018, pp. 9446--9454.

\bibitem{quan2022dual}
Y.~Quan, X.~Qin, T.~Pang, and H.~Ji, ``Dual-domain self-supervised learning and model adaption for deep compressive imaging,'' in \emph{European Conference on Computer Vision}.\hskip 1em plus 0.5em minus 0.4em\relax Springer, 2022, pp. 409--426.

\bibitem{chen2021equivariant}
D.~Chen, J.~Tachella, and M.~E. Davies, ``Equivariant imaging: Learning beyond the range space,'' in \emph{Proceedings of the IEEE/CVF International Conference on Computer Vision}, 2021, pp. 4379--4388.

\bibitem{chen2022robust}
------, ``Robust equivariant imaging: a fully unsupervised framework for learning to image from noisy and partial measurements,'' in \emph{Proceedings of the IEEE/CVF Conference on Computer Vision and Pattern Recognition}, 2022, pp. 5647--5656.

\bibitem{chi2021test}
Z.~Chi, Y.~Wang, Y.~Yu, and J.~Tang, ``Test-time fast adaptation for dynamic scene deblurring via meta-auxiliary learning,'' in \emph{Proceedings of the IEEE/CVF conference on computer vision and pattern recognition}, 2021, pp. 9137--9146.

\bibitem{park2020fast}
S.~Park, J.~Yoo, D.~Cho, J.~Kim, and T.~H. Kim, ``Fast adaptation to super-resolution networks via meta-learning,'' in \emph{European conference on computer vision}.\hskip 1em plus 0.5em minus 0.4em\relax Springer, 2020, pp. 754--769.

\bibitem{wang2024test}
Z.~Wang, Z.~Lu, T.~Wang, Z.~Yang, H.~Yu, Z.~Wang, Y.~Chen, J.~Lu, and Y.~Zhang, ``Test-time adaptation via orthogonal meta-learning for medical imaging,'' \emph{IEEE Transactions on Radiation and Plasma Medical Sciences}, 2024.

\bibitem{qin2023ground}
X.~Qin, Y.~Quan, T.~Pang, and H.~Ji, ``Ground-truth free meta-learning for deep compressive sampling,'' in \emph{Proceedings of the IEEE/CVF Conference on Computer Vision and Pattern Recognition}, 2023, pp. 9947--9956.

\bibitem{darestani2022test}
M.~Z. Darestani, J.~Liu, and R.~Heckel, ``Test-time training can close the natural distribution shift performance gap in deep learning based compressed sensing,'' in \emph{International conference on machine learning}.\hskip 1em plus 0.5em minus 0.4em\relax PMLR, 2022, pp. 4754--4776.

\bibitem{zhao2024test}
Y.~Zhao, T.~Zhang, and H.~Ji, ``Test-time model adaptation for image reconstruction using self-supervised adaptive layers,'' in \emph{European Conference on Computer Vision}.\hskip 1em plus 0.5em minus 0.4em\relax Springer, 2024, pp. 111--128.

\bibitem{afonso2010fast}
M.~V. Afonso, J.~M. Bioucas-Dias, and M.~A. Figueiredo, ``Fast image recovery using variable splitting and constrained optimization,'' \emph{IEEE transactions on image processing}, vol.~19, no.~9, pp. 2345--2356, 2010.

\bibitem{devolder2014first}
O.~Devolder, F.~Glineur, and Y.~Nesterov, ``First-order methods of smooth convex optimization with inexact oracle,'' \emph{Mathematical Programming}, vol. 146, no.~1, pp. 37--75, 2014.

\bibitem{tachella2022unsupervised}
J.~Tachella, D.~Chen, and M.~Davies, ``Unsupervised learning from incomplete measurements for inverse problems,'' \emph{Advances in Neural Information Processing Systems}, vol.~35, pp. 4983--4995, 2022.

\bibitem{xu2024sketched}
G.~Xu, J.~Li, and J.~Tang, ``Sketched equivariant imaging regularization and deep internal learning for inverse problems,'' \emph{arXiv preprint arXiv:2411.05771}, 2024.

\bibitem{zhang2020deep}
K.~Zhang, L.~V. Gool, and R.~Timofte, ``Deep unfolding network for image super-resolution,'' in \emph{Proceedings of the IEEE/CVF conference on computer vision and pattern recognition}, 2020, pp. 3217--3226.

\bibitem{chan2016plug}
S.~H. Chan, X.~Wang, and O.~A. Elgendy, ``Plug-and-play admm for image restoration: Fixed-point convergence and applications,'' \emph{IEEE Transactions on Computational Imaging}, vol.~3, no.~1, pp. 84--98, 2016.

\bibitem{yang2018admm}
Y.~Yang, J.~Sun, H.~Li, and Z.~Xu, ``Admm-csnet: A deep learning approach for image compressive sensing,'' \emph{IEEE transactions on pattern analysis and machine intelligence}, vol.~42, no.~3, pp. 521--538, 2018.

\bibitem{mataev2019deepred}
G.~Mataev, P.~Milanfar, and M.~Elad, ``Deepred: Deep image prior powered by red,'' in \emph{Proceedings of the IEEE/CVF International Conference on Computer Vision Workshops}, 2019, pp. 0--0.

\bibitem{huang2023self}
P.~Huang, C.~Zhang, X.~Zhang, X.~Li, L.~Dong, and L.~Ying, ``Self-supervised deep unrolled reconstruction using regularization by denoising,'' \emph{IEEE transactions on medical imaging}, vol.~43, no.~3, pp. 1203--1213, 2023.

\bibitem{wang2024perspective}
A.~Wang and M.~Davies, ``Perspective-equivariant imaging: an unsupervised framework for multispectral pansharpening,'' \emph{arXiv preprint arXiv:2403.09327}, 2024.

\bibitem{terris2024equivariant}
M.~Terris, T.~Moreau, N.~Pustelnik, and J.~Tachella, ``Equivariant plug-and-play image reconstruction,'' in \emph{Proceedings of the IEEE/CVF Conference on Computer Vision and Pattern Recognition}, 2024, pp. 25\,255--25\,264.

\bibitem{monga2021algorithm}
V.~Monga, Y.~Li, and Y.~C. Eldar, ``Algorithm unrolling: Interpretable, efficient deep learning for signal and image processing,'' \emph{IEEE Signal Processing Magazine}, vol.~38, no.~2, pp. 18--44, 2021.

\bibitem{zhou2024deep}
Q.~Zhou, J.~Qian, J.~Tang, and J.~Li, ``Deep unrolling networks with recurrent momentum acceleration for nonlinear inverse problems,'' \emph{Inverse Problems}, vol.~40, no.~5, p. 055014, 2024.

\bibitem{yang2013linearized}
J.~Yang and X.~Yuan, ``Linearized augmented lagrangian and alternating direction methods for nuclear norm minimization,'' \emph{Mathematics of computation}, vol.~82, no. 281, pp. 301--329, 2013.

\bibitem{robini2018inexact}
M.~C. Robini, F.~Yang, and Y.~Zhu, ``Inexact half-quadratic optimization for linear inverse problems,'' \emph{SIAM Journal on Imaging Sciences}, vol.~11, no.~2, pp. 1078--1133, 2018.

\bibitem{robini2016inexact}
M.~Robini, Y.~Zhu, X.~Lv, and W.~Liu, ``Inexact half-quadratic optimization for image reconstruction,'' in \emph{2016 IEEE International Conference on Image Processing (ICIP)}.\hskip 1em plus 0.5em minus 0.4em\relax IEEE, 2016, pp. 3513--3517.

\bibitem{bolte2014proximal}
J.~Bolte, S.~Sabach, and M.~Teboulle, ``Proximal alternating linearized minimization for nonconvex and nonsmooth problems,'' \emph{Mathematical Programming}, vol. 146, no.~1, pp. 459--494, 2014.

\bibitem{yang2019inexact}
Y.~Yang, M.~Pesavento, Z.-Q. Luo, and B.~Ottersten, ``Inexact block coordinate descent algorithms for nonsmooth nonconvex optimization,'' \emph{IEEE Transactions on Signal Processing}, vol.~68, pp. 947--961, 2019.

\bibitem{nesterov1983method}
Y.~Nesterov, ``A method of solving a convex programming problem with convergence rate o (1/k2),'' in \emph{Soviet Mathematics Doklady}, vol.~27, no.~2, 1983, pp. 372--376.

\bibitem{nesterov2007gradient}
------, ``Gradient methods for minimizing composite objective function,'' UCL, Tech. Rep., 2007.

\bibitem{ehrhardt2024guide}
M.~J. Ehrhardt, Z.~Kereta, J.~Liang, and J.~Tang, ``A guide to stochastic optimisation for large-scale inverse problems,'' \emph{Inverse Problems}, 2025.

\bibitem{tan2024provably}
H.~Y. Tan, S.~Mukherjee, J.~Tang, and C.-B. Sch{\"o}nlieb, ``Provably convergent plug-and-play quasi-newton methods,'' \emph{SIAM Journal on Imaging Sciences}, vol.~17, no.~2, pp. 785--819, 2024.

\bibitem{venkatakrishnan2013plug}
S.~V. Venkatakrishnan, C.~A. Bouman, and B.~Wohlberg, ``Plug-and-play priors for model based reconstruction,'' in \emph{2013 IEEE Global Conference on Signal and Information Processing}.\hskip 1em plus 0.5em minus 0.4em\relax IEEE, 2013, pp. 945--948.

\bibitem{dabov2007image}
K.~Dabov, A.~Foi, V.~Katkovnik, and K.~Egiazarian, ``Image denoising by sparse 3-d transform-domain collaborative filtering.'' \emph{IEEE transactions on image processing: a publication of the IEEE Signal Processing Society}, vol.~16, no.~8, pp. 2080--2095, 2007.

\bibitem{zhang2017beyond}
K.~Zhang, W.~Zuo, Y.~Chen, D.~Meng, and L.~Zhang, ``Beyond a gaussian denoiser: Residual learning of deep cnn for image denoising,'' \emph{IEEE Transactions on Image Processing}, vol.~26, no.~7, pp. 3142--3155, 2017.

\bibitem{chaman2021truly}
A.~Chaman and I.~Dokmani{\'c}, ``Truly shift-equivariant convolutional neural networks with adaptive polyphase upsampling,'' in \emph{2021 55th Asilomar Conference on Signals, Systems, and Computers}.\hskip 1em plus 0.5em minus 0.4em\relax IEEE, 2021, pp. 1113--1120.

\bibitem{ronneberger2015u}
O.~Ronneberger, P.~Fischer, and T.~Brox, ``U-net: Convolutional networks for biomedical image segmentation,'' in \emph{Medical Image Computing and Computer-Assisted Intervention -- MICCAI 2015}, N.~Navab, J.~Hornegger, W.~M. Wells, and A.~F. Frangi, Eds.\hskip 1em plus 0.5em minus 0.4em\relax Cham: Springer International Publishing, 2015, pp. 234--241.

\bibitem{sun2021plug}
Z.~Sun, F.~Latorre, T.~Sanchez, and V.~Cevher, ``A plug-and-play deep image prior,'' in \emph{ICASSP 2021-2021 IEEE International Conference on Acoustics, Speech and Signal Processing (ICASSP)}.\hskip 1em plus 0.5em minus 0.4em\relax IEEE, 2021, pp. 8103--8107.

\bibitem{clark2013cancer}
K.~Clark, B.~Vendt, K.~Smith, J.~Freymann, J.~Kirby, P.~Koppel, S.~Moore, S.~Phillips, D.~Maffitt, M.~Pringle \emph{et~al.}, ``The cancer imaging archive (tcia): maintaining and operating a public information repository,'' \emph{Journal of digital imaging}, vol.~26, pp. 1045--1057, 2013.

\bibitem{huang2015single}
J.-B. Huang, A.~Singh, and N.~Ahuja, ``Single image super-resolution from transformed self-exemplars,'' in \emph{Proceedings of the IEEE conference on computer vision and pattern recognition}, 2015, pp. 5197--5206.

\bibitem{tachella2024unsure}
J.~Tachella, M.~Davies, and L.~Jacques, ``Unsure: Unknown noise level stein's unbiased risk estimator,'' \emph{arXiv preprint arXiv:2409.01985}, 2024.

\end{thebibliography}
